\documentstyle[12pt]{article}

\input epsf
\ifx\epsffile\undefined
\newlength{\epsfysize}
\def\epsffile#1#2#3#4]#5{}
\fi

\renewcommand{\thefootnote}{\fnsymbol{footnote}}
\newcommand{\MS}{{\footnotesize{$\overline{{\rm MS}}$}} }
\def\gtrsim{\raise.3ex\hbox{$>$\kern-.75em\lower1ex\hbox{$\sim$}}}
\def\lesssim{\raise.3ex\hbox{$<$\kern-.75em\lower1ex\hbox{$\sim$}}}

\begin{document}

\begin{flushright}
{\small
SLAC--PUB--7310\\
JHU--TIPAC--96018\\
September 1996\\}
\end{flushright}

\vspace{2cm}

\centerline{WEAK-SCALE PHENOMENOLOGY OF MODELS WITH}
\baselineskip=15pt
\centerline{GAUGE-MEDIATED SUPERSYMMETRY BREAKING}
\baselineskip=32pt

\centerline{\footnotesize JONATHAN A. BAGGER,$^{a}$}
\baselineskip=13pt
\centerline{\footnotesize KONSTANTIN T. MATCHEV,$^{a}$}
\centerline{\footnotesize DAMIEN M. PIERCE\,$^{b}$ and}
\centerline{\footnotesize REN-JIE ZHANG\,$^{a}$}

\baselineskip=22pt

\centerline{\footnotesize\it $^{a}$\,Department of Physics and Astronomy}
\baselineskip=13pt
\centerline{\footnotesize\it Johns Hopkins University}
\centerline{\footnotesize\it Baltimore, Maryland 21218}

\baselineskip=22pt

\centerline{\footnotesize\it $^{b}$\,Stanford Linear Accelerator Center}
\baselineskip=13pt
\centerline{\footnotesize\it Stanford University}
\centerline{\footnotesize\it Stanford, California 94309}

\vspace{1cm}

\abstract{We study in some detail the spectral phenomenology of models
in which supersymmetry is dynamically broken and transmitted to the
supersymmetric partners of the quarks, leptons and gauge bosons, and
the Higgs bosons themselves, via the usual gauge interactions.  We
elucidate the parameter space of what we consider to be the minimal
model, and explore the regions which give rise to consistent radiative
electroweak symmetry breaking.  We include the weak-scale threshold
corrections, and show how they considerably reduce the scale
dependence of the results.  We examine the sensitivity of our results
to unknown higher-order messenger-sector corrections.  We compute the
superpartner spectrum across the entire parameter space, and compare
it to that of the minimal supergravity-inspired model.  We delineate
the regions where the lightest neutralino or tau slepton is the
next-to-lightest supersymmetric particle, and compute the lifetime and
branching ratios of the NLSP.  In contrast to the minimal
supergravity-inspired model, we find that the lightest neutralino can
have a large Higgsino component, of order 50\%.  Nevertheless, the
neutralino branching fraction to the gravitino and the light Higgs
boson remains small, $\lesssim10^{-4}$, so the observation of such a
decay would point to a non-minimal Higgs sector.}

\vspace{2cm}

\vfill

{\noindent\em Work supported by Department of Energy contract
DE--AC03--76SF00515 and by the U.S. National Science Foundation,
grant NSF-PHY-9404057.}

\pagebreak

\normalsize\baselineskip=15pt
\setcounter{footnote}{0}
\renewcommand{\thefootnote}{\arabic{footnote}}

\section{Introduction}

Most studies of supersymmetric phenomenology have focused on models in
which supersymmetry is broken in a hidden sector at a scale of order
$M_{\rm SUSY} \simeq 10^{10}$ GeV.  In these models, supersymmetry
breaking is communicated to the visible sector by gravitational
interactions.  Such models give rise to superpartner masses in the 100
-- 1000 GeV range, and to a gravitino with approximately the same
mass.  The couplings of the gravitino to matter are negligible,
suppressed by $(E/M_{\rm SUSY})^2 \simeq 10^{-16}$.  The now-standard
``minimal supergravity'' model is of this type, with the familiar
unification-scale boundary conditions on $M_0,$ $M_{1/2},$ and $A_0$.

Recently, there has been a resurgence of interest in models where
supersymmetry is broken at a scale of order $M_{\rm SUSY}\gtrsim 10^5$
GeV \cite{DN} -- \cite{Riotto}.  In these models, supersymmetry
breaking is transmitted to the superpartners of the quarks, leptons,
and gauge bosons (and to the Higgs bosons themselves) via the usual
SU(3) $\times$ SU(2) $\times$ U(1) gauge interactions.  Because gauge
interactions are flavor diagonal, these models naturally suppress the
flavor-changing neutral currents associated with the soft squark and
slepton masses.

Models with gauge-mediated symmetry breaking also have superpartner
masses in the 100 -- 1000 GeV range.  However, because of the low
scale of supersymmetry breaking, the gravitino is essentially
massless; its couplings to the superpartners are suppressed by
$(E/M_{\rm SUSY})^2 \lesssim 10^{-6}$.  Typically, superparticles decay
by cascading down to the next-to-lightest supersymmetric particle
(NLSP), which in turn decays to its partner and the gravitino.  Such
decays give rise to characteristic experimental signatures, e.g.
final states containing two photons and missing energy
\cite{pheno,Martin et al}.

In this paper we take a detailed look at the low-energy spectrum of
the simplest models with gauge-mediated supersymmetry breaking.  We
begin by describing the models, elaborating on the issues associated
with electroweak symmetry breaking.  We use two-loop renormalization
group equations and the full one-loop threshold corrections to
determine the parameter space which gives rise to consistent radiative
electroweak symmetry breaking.  We find the Higgs boson and
superpartner masses to the same level of accuracy.  We comment on how
our results are affected by higher-order messenger-sector corrections.
We compare the spectra of the simplest gauge-mediated models with
those from the minimal supergravity-inspired model.  Finally, we
identify the next-to-lightest supersymmetric particle and compute its
lifetime and branching ratios across the allowed parameter space.

\section{Electroweak symmetry breaking}

In models of gauge-mediated supersymmetry breaking, the SU(3) $\times$
SU(2) $\times$ U(1) gauge interactions of ``messenger" fields
communicate supersymmetry breaking from a hidden sector to the fields
of the visible world.  In the simplest models \cite{DN}, the messenger
sector contains a set of vector-like fields, $M_i$ and
$\overline{M}_i$, coupled to a standard-model singlet, $S$, through
the superpotential interaction
\begin{equation}
W_{\rm messenger} = \lambda_i  S M_i \overline{M}_i.
\end{equation}
The lowest and $F$-components of the singlet superfield $S$ acquire
vevs through their interactions with the hidden fields.  This
breaks supersymmetry and $R$-symmetry.

To maintain the near unification of the standard-model gauge
couplings, we will take the fields $M_i$ and $\overline{M}_i$ to
transform in complete SU(5) representations.  In the same spirit, we
will also require that the gauge couplings remain perturbative up to
the unification scale.  This implies that we can consider at most
four $5 + \overline{5}$ pairs, or one $10 + \overline{10}$ and one $5
+ \overline{5}$ pair.

In what follows we will take the fields $M_i$ and $\overline{M}_i$ to
lie in $(n_5,n_{10})$ $5 + \overline{5}$ and $10 + \overline{10}$
SU(5) representations.  We will assume that these fields couple to the
singlet $S$ through a single Yukawa coupling $\lambda$ at the
unification scale.  (See ref.~\cite{Faraggi,Martin} for a discussion of
variations.)  Below that scale, the SU(5) representations split apart,
and the Yukawa couplings evolve according to their own renormalization
group equations \cite{Carone}.  We will ignore this splitting in most
of what follows; we will remark on it briefly in sect.~3.

The lowest and $F$ components of the superfield $S$ acquire vevs
through interactions with the sector which dynamically breaks
supersymmetry. These vevs induce masses and mixings for the messenger
fields.  The messenger fermions gain mass $M=\lambda \langle S
\rangle$, while the messenger scalars obtain the mass matrix
\begin{equation}
\left(\begin{array}{cc}
M^2 & \lambda \langle F_S\rangle\\
\lambda \langle F_S\rangle & M^2
\end{array}\right) .
\end{equation}
{}From this point on, we omit the brackets which denote the vevs.

Following the philosophy of dynamical models, we will assume that the
standard-model $\mu$-term and all soft supersymmetry-breaking terms
arise dynamically.  For the case at hand, the messenger fields
transmit the supersymmetry breaking to the visible sector through loop
diagrams which contain insertions of the $S$ superfield.  Such
diagrams induce weak-scale masses for the gauginos and scalars of
the minimal supersymmetric standard model.  However, they cannot give
sizable values for the soft supersymmetry-breaking $A$ parameters.
In what follows, we shall set all $A$-terms to zero at the messenger
scale.

The messenger fields induce the following gaugino
\begin{equation}
\tilde M_i(M) = (n_5+3n_{10}) g\left({\Lambda\over M}\right)
{\alpha_i(M)\over4\pi} \Lambda
\label{bc1}
\end{equation}
and scalar
\begin{equation}
\tilde m^2(M) = 2  (n_5+3n_{10})
f\left({\Lambda\over M}\right)
\sum_{i=1}^3  k_i   C_i
\biggl({\alpha_i(M)\over4\pi}\biggr)^2
\Lambda^2
\label{bc2}
\end{equation}
masses at the scale $M$, where $\Lambda=F_S/S$ and $k_i=1,1,3/5$ for
SU(3), SU(2), and U(1), respectively.  The $C_i$ are zero for gauge
singlets, and 4/3, 3/4, and $Y^2$ for the fundamental representations
of SU(3), SU(2), and U(1).  Here $Y=Q-I_3$ denotes the
usual hypercharge and we use the grand unification normalization for
$\alpha_1$.  Because a pair of $10 + \overline{10}$ fields contributes
to the soft masses as if $n_5=3$, we will set $n_{10}=0$ and only consider
changes in $n_5$.

The messenger-scale threshold functions \cite{Martin,Pomarol}
\begin{equation}
g(x)={1+x\over x^2}\log(1+x)  +   (x\rightarrow-x)\ ,
\end{equation}
\begin{equation}
f(x)={1+x\over x^2}\biggl[\log(1+x) -2{\rm Li}_2
\left({x\over1+x}\right)+\ {1\over2}{\rm Li}_2
\left({2x\over1+x}\right)\biggr]   +   (x\rightarrow-x)\ ,
\end{equation}
have the property that $g(x), f(x)\rightarrow1$ as $x\rightarrow0$,
or $\Lambda \ll M.$ In this limit, the expressions (\ref{bc1}) and
(\ref{bc2}) take the characteristic simple forms \cite{DN} that are
often associated with gauge-mediated models.

The region $\Lambda \rightarrow M$ corresponds to $x \rightarrow 1$,
where $g(1) \simeq 1.4$ and $f(1) \simeq 0.7$.  We must exclude the
limit $M=\Lambda$ because it gives rise to a massless messenger scalar.
For the purposes of this paper, we restrict ourselves to the region
$x<0.97$, or $M/\Lambda>1.03$.  This corresponds to an upper limit on
the fine tuning of the messenger masses; it is obtained by requiring
that the average scalar mass-squared be less than 30 times the light
scalar mass-squared.

Equations (\ref{bc1}) and (\ref{bc2}) serve as boundary conditions
for the renormalization group equations at the messenger scale,
$M$.  They give rise to rather generic predictions for the soft
supersymmetry-breaking gaugino and scalar masses.  In contrast,
the boundary conditions for $B(M)$ and $\mu(M)$ are more
model-dependent.\footnote{The parameter $\mu$ is the standard
supersymmetric Higgsino mass; $B$ is the dimension-two soft mass
that is often denoted $B\mu$ in supergravity-inspired models.}
New interactions are necessary to induce these terms because
they violate a Peccei-Quinn symmetry in the effective action
\cite{mu-term,DNS}.  The new interactions can give rise to additional
contributions to the scalar masses beyond those in (\ref{bc2}).  In
particular, they can give additional contributions to the Higgs
masses, $m_{H_1}^2$ and $m_{H_2}^2$.

In this paper, we will not commit ourselves to a particular model for
$B(M)$ and $\mu(M)$.  Instead, we will take a more phenomenological
approach and treat them as free parameters.  We will, however, assume
that the soft Higgs masses $m_{H_1}^2$ and $m_{H_2}^2$ are given by
eq.~(\ref{bc2}).  We will also require that electroweak symmetry be
radiatively broken.

Our approach is as follows.  We start at the messenger scale, $M$, and
fix the boundary conditions (\ref{bc1}) and (\ref{bc2}).  We use the
two-loop renormalization group equations to run the soft masses down
to the squark mass scale, $M_{\tilde q}\simeq \sqrt{n_5}
 \Lambda/90$, where we impose electroweak symmetry breaking and
calculate the supersymmetric mass spectrum.  At the squark scale, we
consistently include all one-loop weak-scale threshold corrections.
These corrections play an important role in determining $B(M_{\tilde
q})$ and $\mu(M_{\tilde q})$.  We then run $B$, $\mu$ and the gauge
and Yukawa couplings back to the messenger scale.  We repeat the
procedure until we find a self-consistent solution for $B(M)$ and
$\mu(M)$ in terms of the $Z$-boson pole mass, $M_Z$, and the ratio of
Higgs vacuum expectation values, $\tan\beta$.  (We take the top- and
bottom-quark pole masses to be $m_t = 175$ GeV, $m_b = 4.9$ GeV, and
the \MS value of the strong coupling to be $\alpha_s(M_Z) = 0.118.$)

We note that there are messenger-scale threshold corrections which
we do not take into account, even though they are formally of the
same order as the weak-scale threshold corrections.  We choose
to ignore these corrections because they are model dependent.  We
will, however, estimate their importance by determining the
sensitivity of our results to small changes in the messenger-scale
boundary conditions.

With these assumptions, the parameter space of this minimal model is
described by $\tan\beta$, the scale $\Lambda$, the messenger scale
$M$, along with $n_5$, the effective number of $5+\overline{5}$
messenger fields, and the sign of $\mu$.  We will first set $n_5=1$
and $M=2\Lambda$.  This leaves a two-dimensional parameter space,
$(\Lambda,\tan\beta)$, for each sign of $\mu$.

\begin{figure}[t]
\epsfysize=2.5in
\epsffile[-180 180 -40 545]{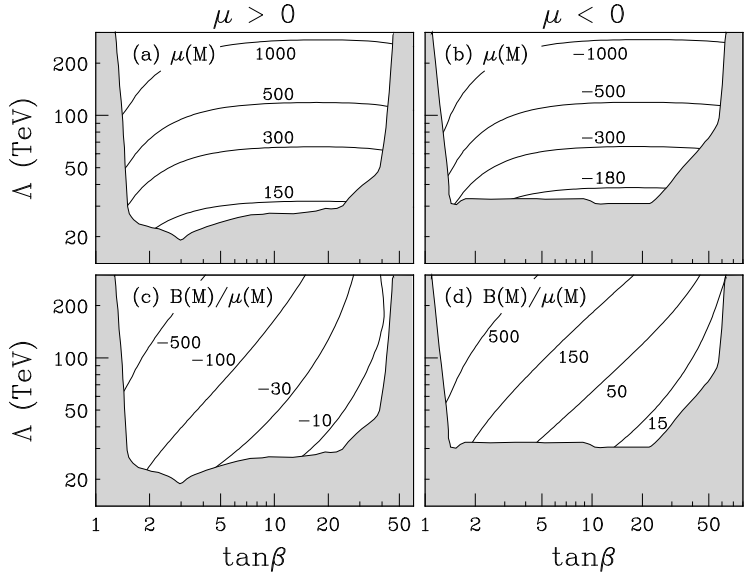}
\begin{center}
\parbox{5.5in}{
\caption[]{\small Contours of $\mu(M)$ and $B(M)/\mu(M)$ in the
$\Lambda, \tan\beta$ plane, with $n_5=1$ and $M/\Lambda=2$.
The contours are labeled in GeV.
\label{f.Bmu}}}
\end{center}
\end{figure}

In Fig.~\ref{f.Bmu} we show our results for $B(M)/\mu(M)$ and $\mu(M)$
in the $(\Lambda, \tan\beta)$ plane.  We see that consistent
electroweak symmetry breaking requires $|\mu(M)|$ to be in the range
from 150 GeV for small $\Lambda$ to over 1 TeV for large $\Lambda$.  We
also find that $B(M)/\mu(M)$ ranges from near zero, at the largest
values of $\tan\beta$, to $ \gtrsim 500$ GeV at small $\tan\beta$.  We
do not consider $\Lambda>300$ TeV because of fine tuning
considerations.

Various constraints exclude the border regions in Fig.~\ref{f.Bmu}
(and the remaining contour plots).  The region of small $\tan\beta$ is
excluded because the top Yukawa coupling diverges below the
unification scale.  This is not necessarily fatal, but since we wish
to preserve perturbative grand unification, we require
$\lambda_t(M_{\rm GUT}) < 3.5$.  The region of very large $\tan\beta$
is excluded because electroweak symmetry is not broken.  Typically, we
find $m_A^2<0$ in the large $\tan\beta$ region.

The region at small $\Lambda$ is excluded because of the usual
experimental bounds on the masses of supersymmetric particles, most
notably $m_{\tilde\chi^+} > 65$ GeV.  (We use the bounds listed in
Ref.~\cite{PBMZ}.)  These bounds hold when the LSP is the lightest
neutralino and is stable.  They also hold in gauge-mediated models
when the NLSP is $\tilde\chi_1^0$ and it decays outside the detector.
In sect.~4 we delineate the regions of the parameter space where this
applies.  If the NLSP decays inside the detector, the collider bounds
are stronger because the decay modes
(e.g. $\tilde\chi_1^0\tilde\chi_1^0\rightarrow \gamma\gamma +$
missing energy) have much smaller backgrounds \cite{pheno}.

\begin{figure}[t]
\epsfysize=2.5in
\epsffile[-140 215 0 540]{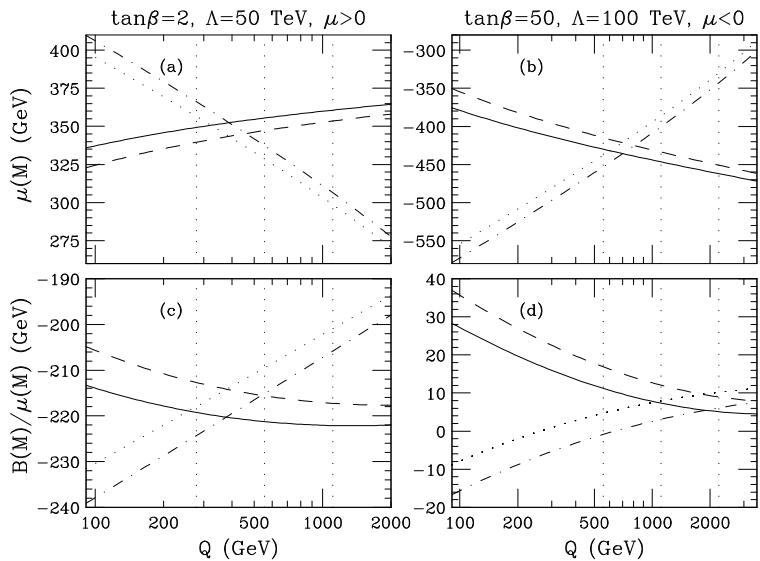}
\begin{center}
\parbox{5.5in}{
\caption[]{The parameters $\mu(M)$ and $B(M)/\mu(M)$
as a function of the scale, $Q$, where the renormalization
group evolution is stopped and electroweak symmetry breaking is
imposed.  The dotted (dot-dashed) lines include the one-loop
(two-loop) evolution equations, but no weak-scale thresholds.  The
dashed (solid) lines include the one-loop (two-loop) evolution
equations, plus the full set of one-loop weak-scale thresholds.  The
dotted vertical lines on each plot denote the squark scales
$M_{\tilde q}/2$, $M_{\tilde q}$ and $2M_{\tilde q}$.
\label{f.scale}}}
\end{center}
\end{figure}

In Fig.~\ref{f.scale} we plot the scale dependence of $B(M)/\mu(M)$
and $\mu(M)$ to illustrate the importance of the weak-scale threshold
corrections.  We see that our results, which consistently incorporate
the weak-scale thresholds, have significantly less scale dependence
than they would if the corrections were ignored.  Furthermore, we see
that the threshold corrections are especially important at large
$\tan\beta$, where $B(M)$ is small.  The largest threshold corrections
arise from squark loops, so the threshold corrections are smallest in
the vicinity of the squark scale, $M_{\tilde q} = \sqrt{n_5}
\Lambda/90$.  Indeed, in Fig.~\ref{f.scale} we see that the
tree-level result is nearly equal to the full result at the scale
$M_{\tilde q}/2$.  If we vary the scale over a reasonable range,
e.g. $M_{\tilde q}/2$ to $2M_{\tilde q}$, we find that the weak-scale
threshold corrections (between $5$\% and $15$\%) are generally larger
than the corrections from the two-loop evolution (about $2\%$), as
might be expected because of the relatively small amount of running.

The region $B(M) \simeq 0$ is of considerable phenomenological
interest \cite{DNS,BKW}.  For instance, models where $B(M) \simeq 0$
give rise to large $\tan\beta$ without fine tuning \cite{DNS}.
Furthermore, since the $A$-terms are also small at the scale $M$, such
models have a naturally small neutron electric dipole moment
\cite{DNS}.  Indeed, only the small ${\cal O}(\alpha M_2/\pi)$
starting values of $A(M)$ and $B(M)/\mu(M)$ can contribute to the
neutron EDM.  The pieces of $A$ and $B$ which are generated by the
renormalization group, $A_{RG}$ and $B_{RG}$, do not contribute
because they are proportional to $M_2$ and $\mu M_2$, so the invariant
CP-violating phases \cite{CP} arg($A_{RG}^*M_2$) and arg($B_{RG}^*\mu
M_2$) vanish.

The region of $B(M) \simeq 0$ is associated with the region
of large $\tan\beta$.  At tree level, we have
\begin{equation}
B = {m_{H_1}^2 - m_{H_2}^2 \over \tan\beta-\cot\beta}
 -  {M_Z^2\over\tan\beta+\cot\beta} ,
\end{equation}
which implies that $B$ decreases as $\tan\beta$ increases.  From
Fig.~\ref{f.Bmu} we find, for $n_5=1$ and $M=2\Lambda$, that $B(M)$ is
small (but non-zero) along the excluded region at large $\tan\beta$.
(Our results should be contrasted with those of ref.~\cite{BKW}, which
do not include one-loop thresholds.  The authors of ref.~\cite{BKW}
find $B(M)=0$ for $\tan\beta \simeq 20$ and $\mu < 0$.)

Until now we have restricted our attention to the simple case where
$M/\Lambda =2.$  In principle, $M/\Lambda$ can be much larger.  A
large hierarchy $M\gg\Lambda$ can arise from a small Yukawa coupling
or from loop factors \cite{Hotta}.  An upper bound on this splitting
can be obtained from the cosmological constraint that the gravitino
relic density not overclose the universe.  In the usual cosmological
scenario, this translates into an upper bound on the gravitino mass
of about 1 keV \cite{grav bound}.  (Note, however, that larger masses
can be accommodated if the gravitino relic density is suppressed
by a brief inflationary epoch.)  If we assume that the singlet
$F$-term is of the same order of magnitude as the largest $F$-term
in the theory, we have
\begin{equation}
m_{\tilde G} \simeq {F_S \over M_P} = \lambda^{-1}
\left({M\over\Lambda}\right){\Lambda^2\over M_P},
\end{equation}
Setting $\Lambda=30$ TeV and $\lambda=1$, we find that
$m_{\tilde G} \simeq 1$ keV
if $M/\Lambda\simeq10^4$. Hence we will not consider values of
$M/\Lambda$ larger than $10^4$. Note that for $\lambda < 1$
the upper bound on $M/\Lambda$ is correspondingly reduced.

\begin{figure}[t]
\epsfysize=2.5in
\epsffile[-180 180 -40 545]{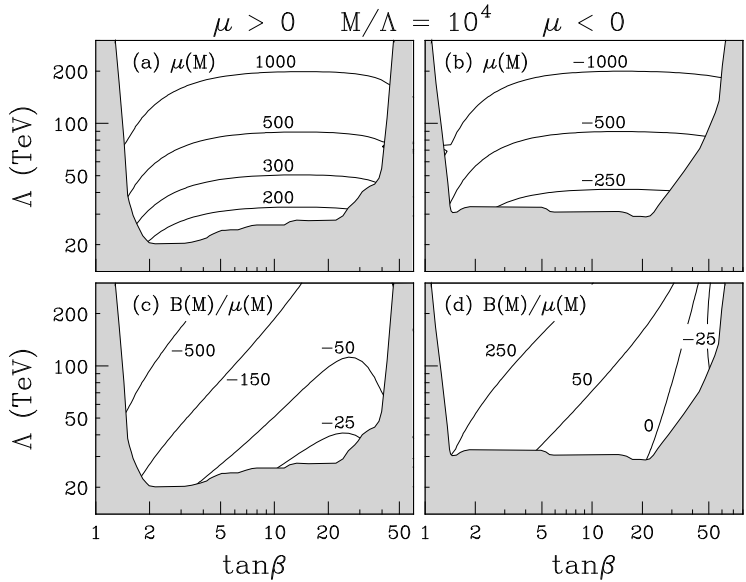}
\begin{center}
\parbox{5.5in}{
\caption[]{The same as Fig.~\ref{f.Bmu}, with $M/\Lambda = 10^4$ and
$n_5 = 1$.  The magnitude of $\mu(M)$ is increased relative to that of
Fig.~1.
\label{f.MoL}}}
\end{center}
\end{figure}

As we increase $M$ for a fixed $\Lambda$, there is more renormalization
group evolution, so there are larger splittings between the soft Higgs
masses.  These, in turn, give somewhat larger values of $\mu(M)$.
We show this in Figs.~\ref{f.MoL}(a-b), where we plot
contours of $\mu(M)$ in the $(\Lambda, \tan\beta)$ plane, for $n_5=1$
and $M/\Lambda =10^4$.  We see that the values of $\mu(M)$ are
slightly larger than those in Fig.~\ref{f.Bmu}.  The situation for
$B(M)$ is more subtle because the region $B(M) \simeq 0$ is sensitive
to additive radiative corrections.  In Figs.~\ref{f.MoL}(c-d) we show
our results for $M/\Lambda=10^4$.  We find that $B(M)=0$ occurs for
$\mu<0$ with $\tan\beta \simeq 20$ to 40, depending on $\Lambda$. For
smaller $M/\Lambda$, the $B(M)=0$ contour in Fig.~\ref{f.MoL}(d) moves
to the right, to larger values of $\tan\beta$.

\begin{figure}[t]
\epsfysize=2.5in
\epsffile[-180 180 -40 545]{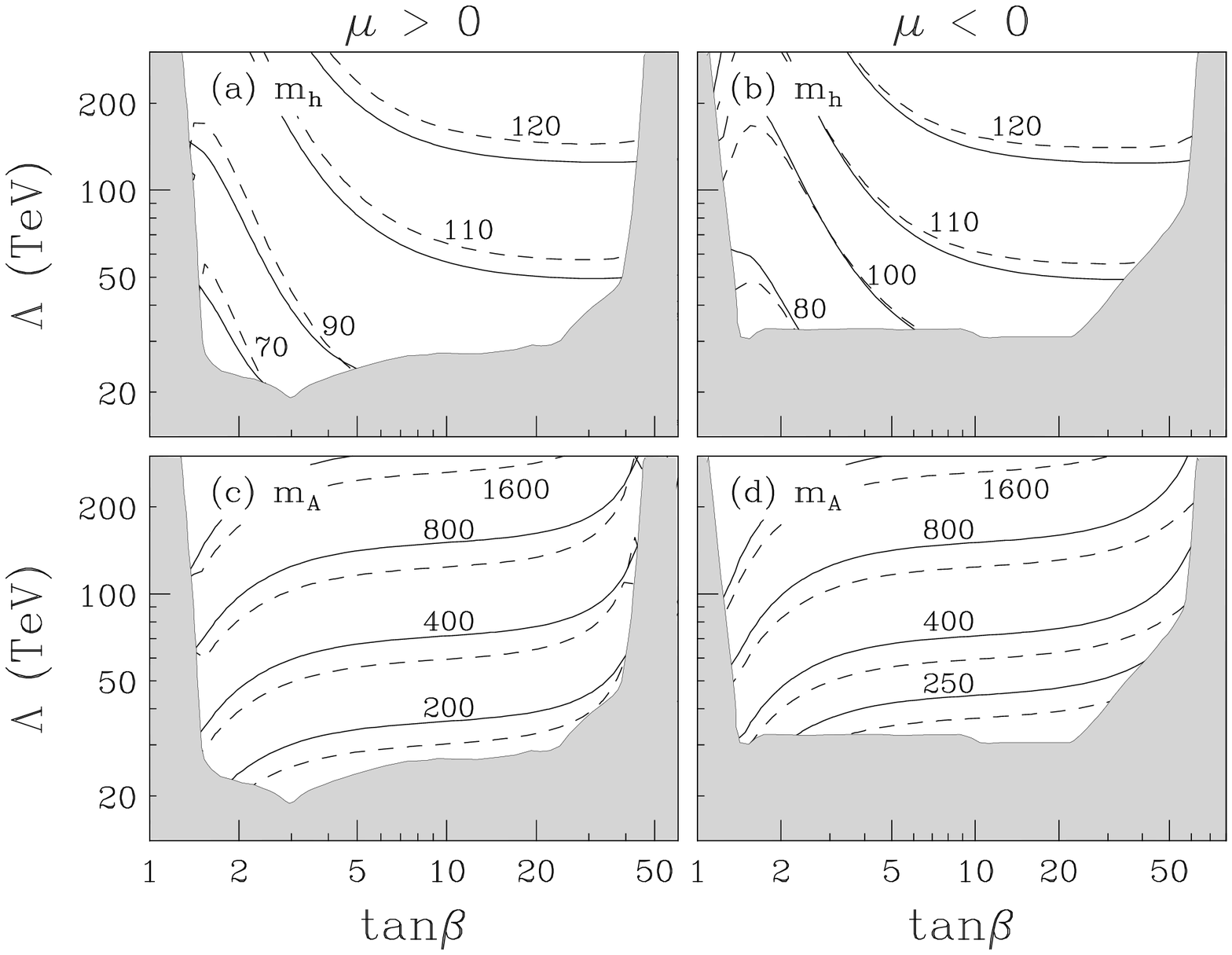}
\begin{center}
\parbox{5.5in}{
\caption[]{Figures (a) and (b) show contours of the light CP-even
Higgs-boson mass, $m_h$, in GeV, while Figs.~(c) and (d) show the
CP-odd Higgs-boson mass, $m_A$, in GeV.  We set $n_5 = 1$, and
the solid (dashed) lines correspond to $M/\Lambda = 2 (10^4)$.
\label{f.higgs}}}
\end{center}
\end{figure}

In Fig.~\ref{f.higgs} we plot the pole mass of the lightest Higgs
boson, $m_h$, and of the CP-odd Higgs boson, $m_A$, in the $(\Lambda,
\tan\beta)$ plane, for $M/\Lambda = 2$ and $10^4$.  We see that $m_h
\lesssim 130$ GeV (see, however, Ref.~\cite{Riotto}),
and that $m_A$ is nearly always larger than 200 GeV.
For such large values of $m_A$, all three heavy Higgs bosons are
nearly degenerate in mass.

\begin{figure}[t]
\epsfysize=2.5in
\epsffile[-180 180 -40 545]{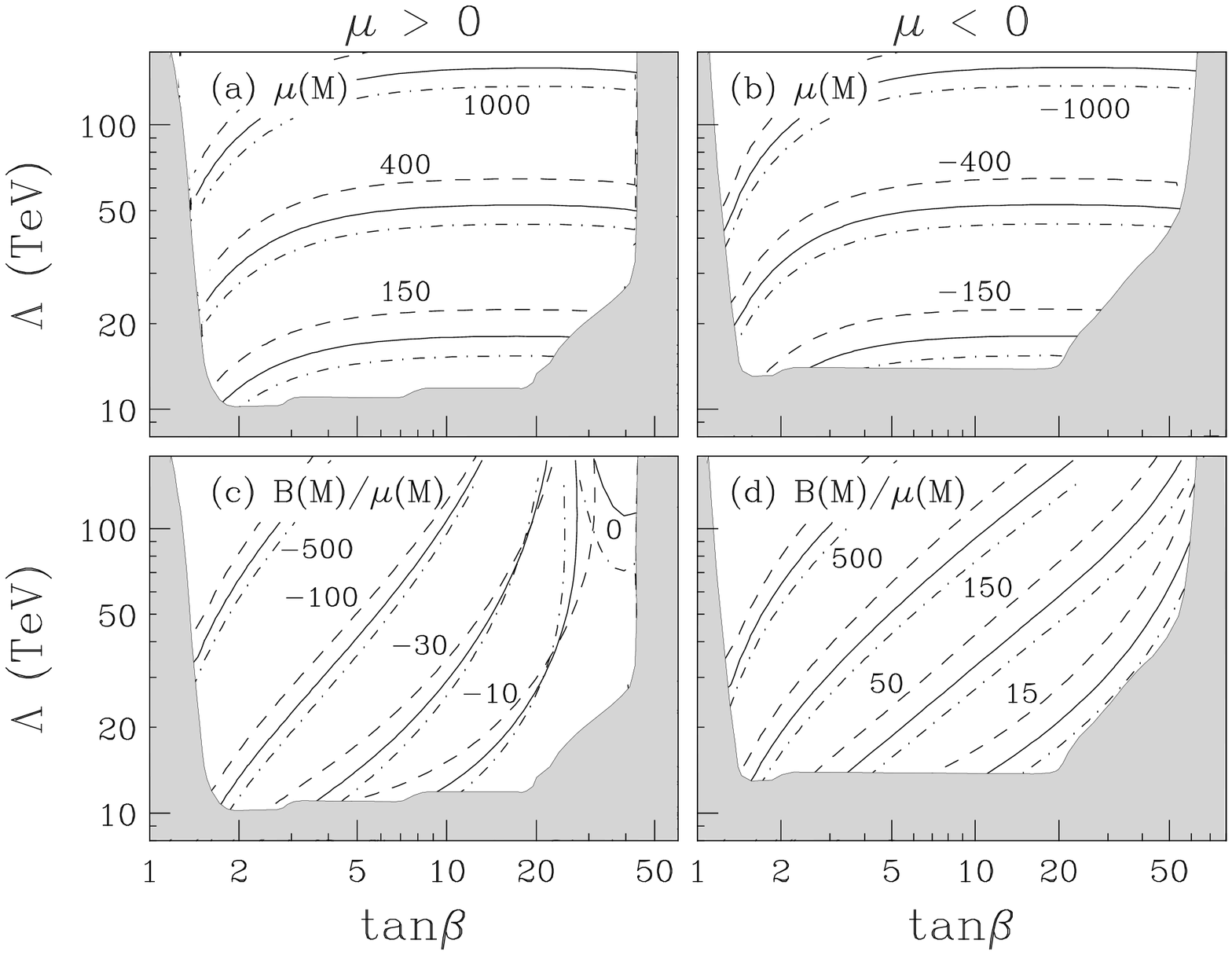}
\begin{center}
\parbox{5.5in}{
\caption[]{The same as Fig.~\ref{f.Bmu} for varying $n_5$.  The
dashed, solid and dot-dashed lines correspond to $n_5 = 2, 3, 4$,
respectively.  The values of $\mu(M)$ and $B(M)/\mu(M)$ scale like
$\sqrt n_5$ for a fixed value of $\Lambda$, except for $B(M) \simeq
0$, where radiative corrections are important.
\label{f.n5}}}
\end{center}
\end{figure}

The above results depend sensitively on $n_5$, the number of $5 +
\overline{5}$ pairs.  In Fig.~\ref{f.n5} we show, for $M=2\Lambda$,
contour plots with $n_5 = 2, 3$ and $4$.  The change in the
parameter $\mu(M)$ is dominated by the change in the scalar masses.
Indeed, we find $\mu(M)/\Lambda \propto \sqrt{n_5}$, as expected from
eq.~(\ref{bc2}).  The ratio $B(M)/\mu(M)$ obeys the same scaling at
small $\tan\beta$.  At large $\tan \beta$, however, the change in
$B(M)$ is more complicated.  This is the region where $B(M)$ is small,
so its value depends sensitively on radiative corrections.  For $\mu >
0$, there is a line in parameter space where $B(M) = 0$ at large
$\Lambda$. For $\mu < 0$, however, $B(M)$ is never zero.  We find
$B(M)/\mu(M) > 10$ GeV for $n_5 = 4$.

The results presented here have varying degrees of sensitivity to the
input parameters.  We illustrate this sensitivity by computing the
changes in $B(M)/\mu(M)$ and $\mu(M)$.  We first vary the quark
masses, $m_b=4.9 \pm 0.5$ GeV and $m_t=175 \pm 5$ GeV, and find a
shift $\Delta|B/\mu| \lesssim 10$ GeV and $|\Delta\mu/\mu| \lesssim 7\%$ in
the region $B(M) \simeq 0$.  To estimate the sensitivity to
messenger-scale threshold corrections, we randomly vary the soft
masses at the messenger scale by $5\%$.  In the region $B(M) \simeq
0$, we find shifts in $B(M)/\mu(M)$ of up to $10$ GeV, and changes in
$\mu$ of order $5$\%.

\section{The spectrum}

In this section we find the masses of the supersymmetric particles
that arise in the gauge-mediated scheme.  We examine how the
spectrum depends on the parameter space, and comment on the sensitivity
to low- and high-energy threshold corrections.  As above, we follow a
self-consistent procedure.  We start with the masses (\ref{bc1}) and
(\ref{bc2}) at the messenger scale, $M$.  We then use the two-loop
renormalization group equations to run these masses to the squark
scale, $M_{\tilde q}=\sqrt{n_5} \Lambda/90$.  At that scale we apply
the one-loop threshold corrections and impose electroweak symmetry
breaking.  We then calculate the superpartner masses, and run the
gauge couplings back to the messenger scale.  We iterate this
procedure to determine the consistent one-loop superpartner pole
masses.  The weak-scale threshold corrections for all the superpartner
masses are contained in ref.~\cite{PBMZ}.

\begin{figure}[t]
\epsfysize=2.4in
\epsffile[-15 320 125 550]{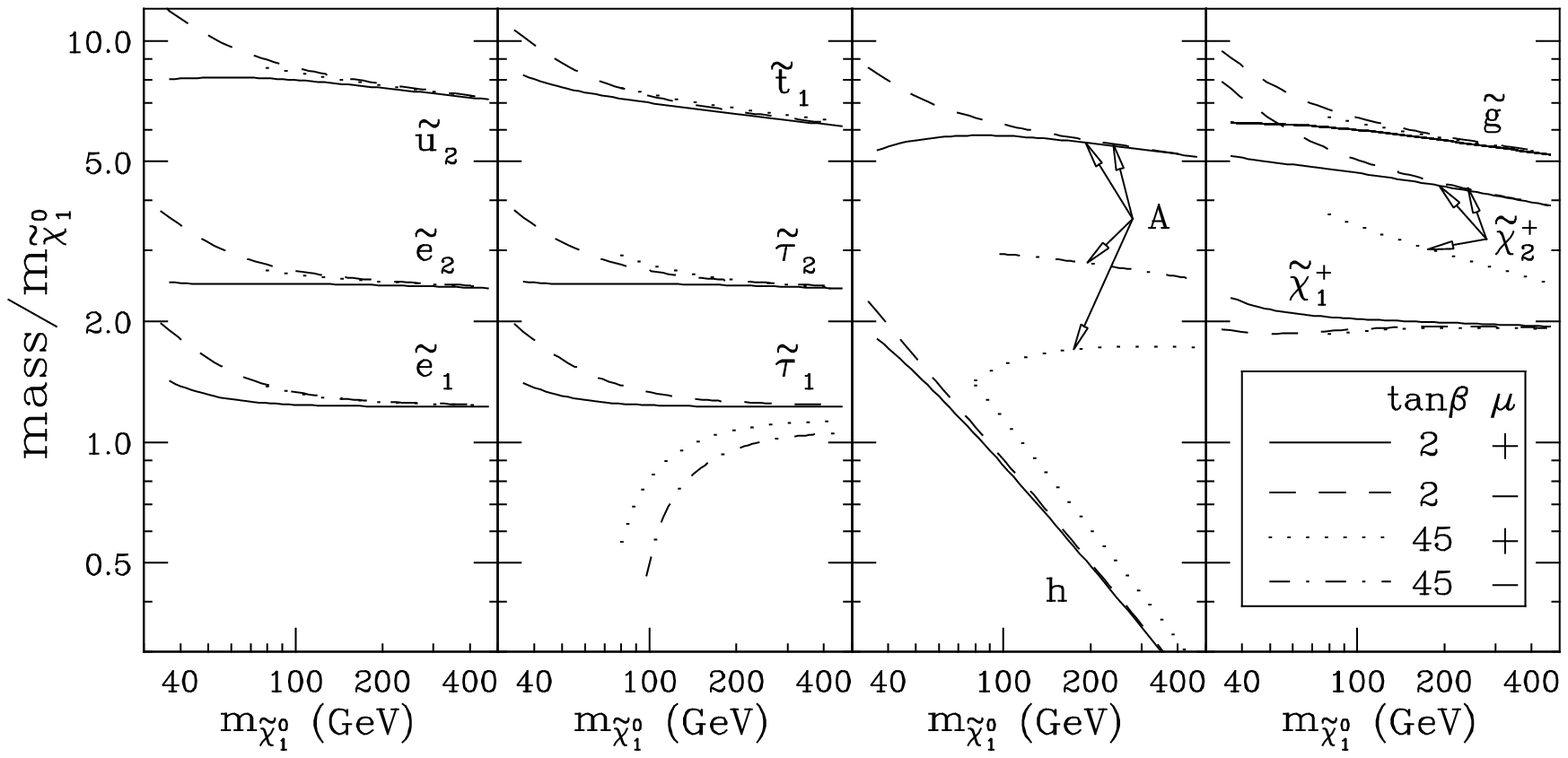}
\begin{center}
\parbox{5.5in}{
\caption[]{The ratio of various superpartner masses to
$m_{\tilde\chi_1^0}$, versus $m_{\tilde\chi_1^0}$, for $M=2\Lambda$
and $n_5=1$.  The different lines for each particle correspond to the
cases of small/large $\tan\beta$ and positive/negative $\mu$.
Except for $m_A$ and $m_{\tau_1}$, the large $\tan\beta$ curves are
essentially independent of the sign of $\mu$.
\label{f.spec}}}
\end{center}
\end{figure}

Figure \ref{f.spec} shows the spectrum in the canonical case where the
messenger scale $M$ is equal to $2 \Lambda$.  We plot the various
masses against the mass of the lightest neutralino, $\tilde\chi_1^0$.
The first point to note is that, for $m_{\tilde\chi_1^0} \gtrsim 100$
GeV, most of the curves are rather flat.  This simply reflects the
fact that most of the masses, including $m_{\tilde\chi_1^0}$, scale
with $\Lambda$.  The only exceptions are the light Higgs, $h$, whose
mass is determined by radiative corrections (the correction to its
mass-squared grows like $\ln\Lambda$), and the light tau slepton,
whose mass is significantly affected by left-right mixing.

\begin{figure}[t]
\epsfysize=1.8in
\epsffile[-10 378 0 548]{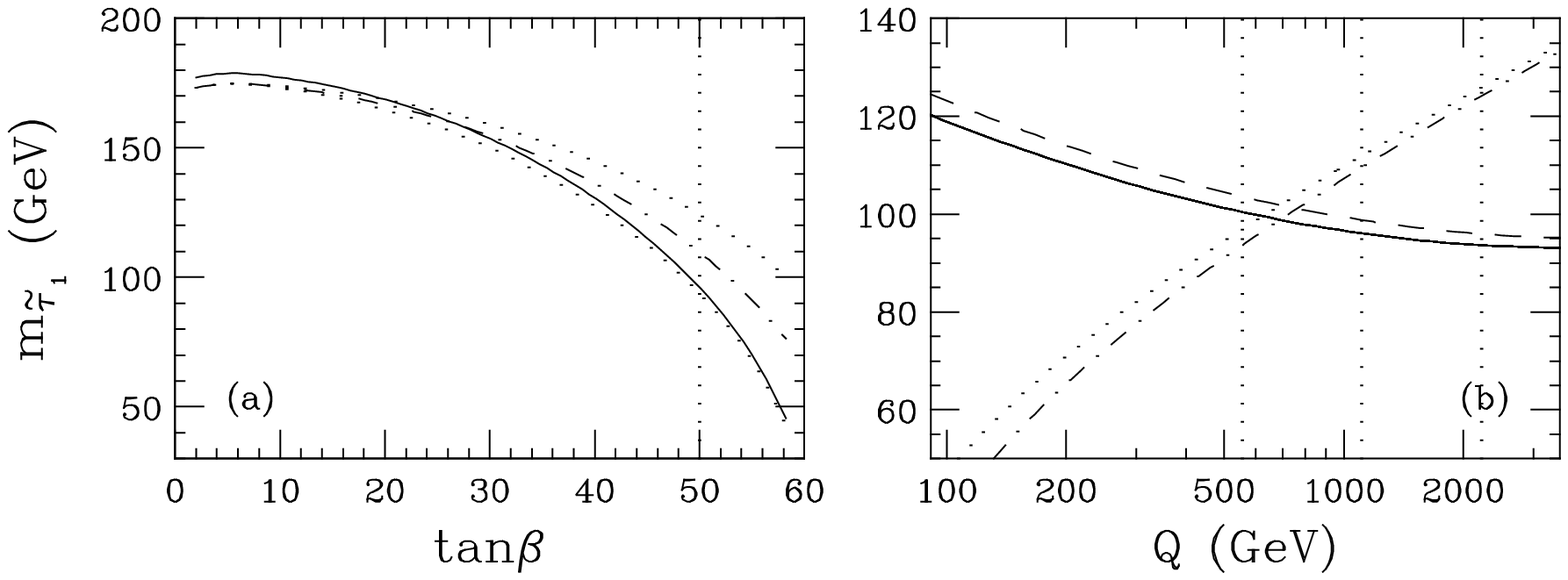}
\begin{center}
\parbox{5.5in}{
\caption[]{(a) The light tau slepton mass versus
$\tan\beta$, for $n_5=1$, $\Lambda=100$ TeV, $M=2\Lambda$, and
$\mu<0$. The one-loop result (solid) and tree-level results for
$Q=M_{\tilde q}/2,$ $M_{\tilde q},$ and $2M_{\tilde q}$ are shown.
The tree-level curve for $Q=M_{\tilde q}/2$ is closest to the full
result.
(b) The scale dependence of the tau slepton mass for
the same parameters, with $\tan\beta=50$. The full result (solid)
is contrasted with the tree-level result (dot-dashed).  The
corresponding curves obtained using one-loop renormalization
group evolution are shown (dashed, dotted).
\label{f.stau}}}
\end{center}
\end{figure}

Because of the large mixing in the tau slepton mass matrix, the
light tau slepton mass is especially sensitive to radiative corrections.
The one-loop self-energy correction is typically less than ${\cal O}
(3\%)$ (see Ref.~\cite{PBMZ}).  However, for large $\tan\beta$, the
large threshold corrections to $\mu$ (see Fig.~\ref{f.scale}) induce a
significant shift in the light tau slepton mass.  In
Fig.~\ref{f.stau}(a) we plot the light tau slepton mass versus
$\tan\beta$.  We show
the one-loop and tree-level results, for various choices of the
renormalization scale, $Q$.  At large $\tan\beta$, the corrections can
be of order 30\%.  In Fig.~\ref{f.stau}(b) we show the scale
dependence of the one-loop and tree-level tau slepton masses, as well
as the effect of the two-loop renormalization group evolution.
The weak-scale threshold corrections significantly reduce
the scale dependence of the pole mass. As expected, the two-loop
renormalization group effects are small because of the small amount
of running.

The tau slepton mass matrix should be contrasted with those
of the top and bottom squarks.  These matrices can also have
substantial mixing, but the mixing does not significantly change the
mass eigenvalues.  This implies that the masses of the light top
and bottom squarks cannot be much less than the masses of the other
squarks.  Indeed, we find that the mass of the light top squark is
typically 10 to 20\% less than that of the other squarks, although it
can be as much as 35\% less at the smallest values of $\tan\beta$.
Similarly, the light bottom squark mass is usually 0 to 10\% less
than that of the other squarks, but it can be as much as 25\% less at
the largest values of $\tan\beta$.

\begin{figure}[t]
\epsfysize=2in
\epsffile[-180 240 -40 490]{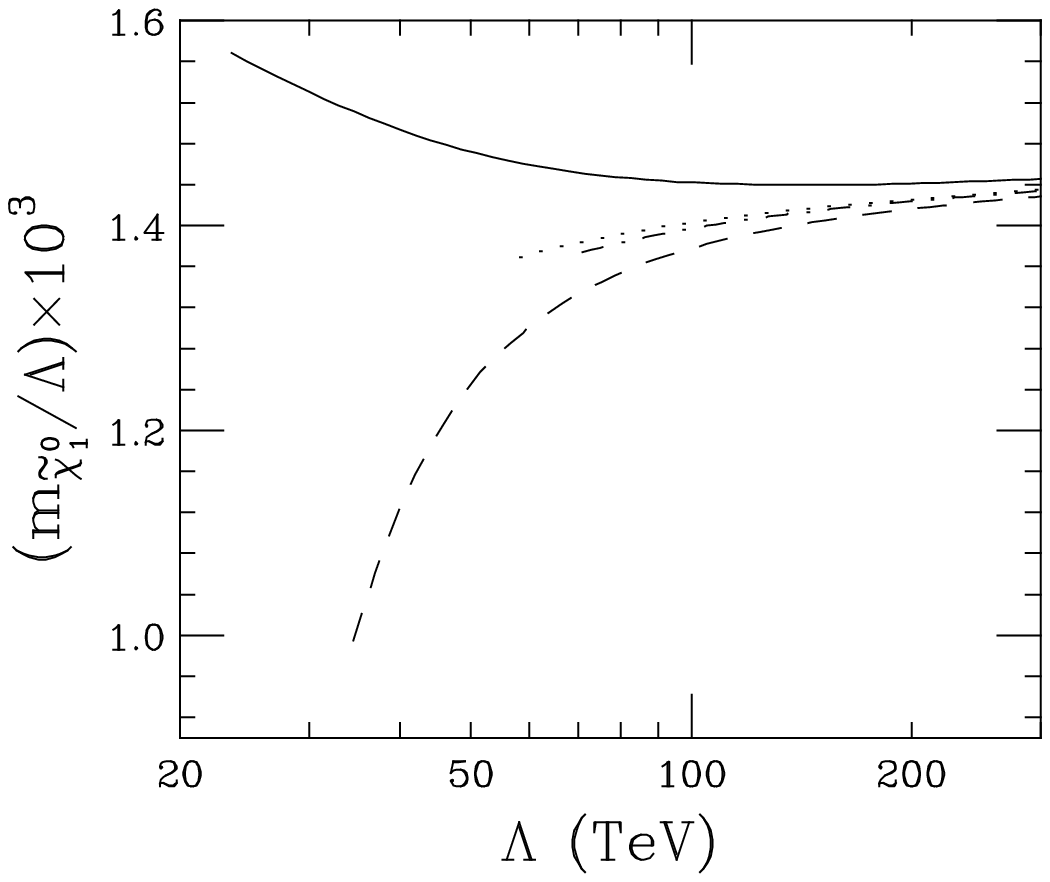}
\begin{center}
\parbox{5.5in}{
\caption[]{The ratio $m_{\tilde\chi_1^0}/\Lambda$ versus $\Lambda$,
for the same choice of parameters as in Fig.~\ref{f.spec}.
\label{f.mchi0}}}
\end{center}
\end{figure}

In Fig.~\ref{f.spec} we see that for $m_{\tilde\chi_1^0} \lesssim 100$
GeV, the curves leave their linear trajectories, especially for
$\tan\beta =2$ with $\mu$ negative.  This is because for small
$m_{\tilde\chi_1^0}$, the mixing in the neutralino mass matrix become
increasingly important, so the $\tilde\chi_1^0$ mass no longer scales
with $\Lambda$.  This is illustrated in Fig.~\ref{f.mchi0}, where we
show the $\tilde\chi_1^0$ mass as a function of $\Lambda$.

The previous spectra were all computed with $M = 2 \Lambda$.  If $M$
is increased with respect to $\Lambda$, the masses change in two ways.
First, the initial conditions are different because the masses at the
scale $M$ depend on $\alpha_i(M)$.  Second, more running is needed to
reach the squark scale, $M_{\tilde q}$.

\begin{figure}[t]
\epsfysize=2.4in
\epsffile[-20 320 120 550]{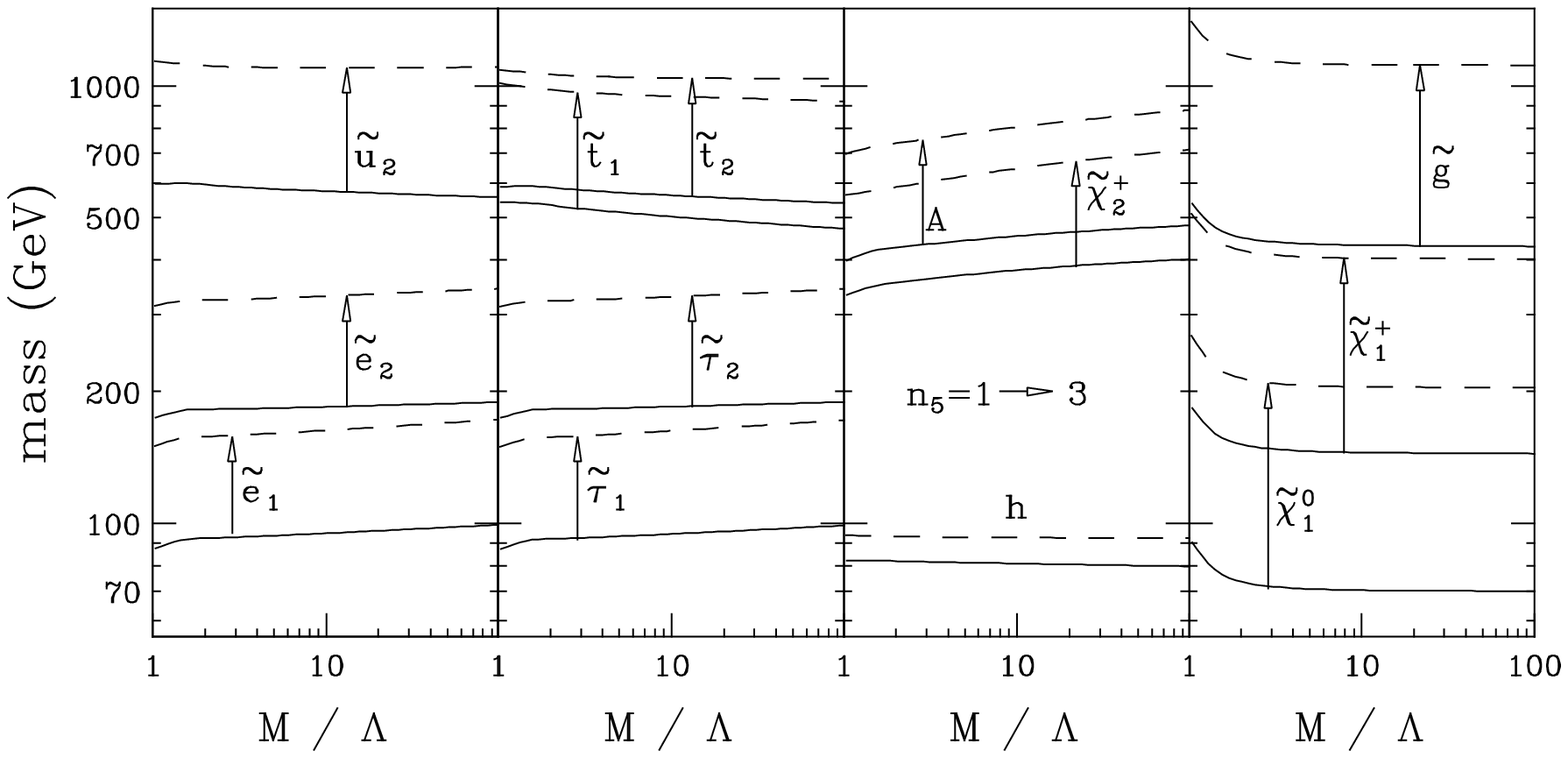}
\begin{center}
\parbox{5.5in}{
\caption[]{Various sparticle masses versus the ratio $M/\Lambda$, for
$\Lambda=50$ TeV, $\tan\beta=2$ and $\mu>0$.  The solid (dashed) lines
correspond to $n_5=1$ ($n_5=3$).
\label{f.spec.MoL}}}
\end{center}
\end{figure}

These effects almost cancel for the gaugino masses because the gaugino
masses obey the same one-loop renormalization group equations as the
gauge couplings.  They do, however, change the scalar masses.  For
squarks, the two effects work in opposite directions.  It turns out
that for $n_5=1$ the change in the boundary conditions is more
important, so the squark masses are reduced.  For $n_5=3$ the two
effects largely cancel.  For sleptons, the effects work in the same
direction, and give a slight increase in the slepton masses.  The
shifts in the scalar masses are illustrated in Fig.~\ref{f.spec.MoL}.
The small changes indicate that it is safe to ignore any factor of two
or three change in the messenger quarks scale with respect to the
messenger lepton scale, such as might be induced by renormalization
group evolution of the messenger Yukawa coupling $\lambda$.

Figure \ref{f.spec.MoL} also illustrates the importance of the
threshold functions $f(x)$ and $g(x)$.  In the region $1.03 \lesssim
M/\Lambda \lesssim 2$ the threshold functions lead to significant
increases in the gaugino masses, and to small decreases in the scalar
masses, relative to the case when $M/\Lambda \gtrsim 2$.  The threshold
corrections are so important that the light tau slepton becomes the
NLSP for $M$ near $\Lambda$.

In Fig.~\ref{f.spec.MoL} we also show the effect of increasing $n_5$
from 1 to 3.  The gaugino masses scale like $n_5$, while
the scalar masses go like $\sqrt n_5$.  Because of this fact, for the
case of a $10+ \overline{10}$ pair of messenger fields ($n_5=3$), the
light tau slepton is the NLSP over most of the parameter space.  We
discuss this in more detail in the next section.

Note that the squark masses are light enough that they will be
produced in the next generation of colliders over most of the
parameter space.

As above, we can estimate the sensitivity of the supersymmetric spectrum
to messenger-scale thresholds by varying the soft masses at the
messenger scale by $5\%$.  We find that most of the pole masses vary
by about $5$\%, except for the light Higgs, which varies by $\lesssim
1$\%, and the top and bottom squarks, which change by up to $
10$\%.  The light tau slepton mass varies by up to $15$\%, except
at a few exceptional points, where the variation can be as large as
70\%.  This large sensitivity in the tau slepton mass occurs at large
$\tan\beta$, in the region where it is potentially the NLSP.

We will conclude this section by contrasting the predictions of the
gauge-mediated models with the predictions of the minimal supergravity
model.  At the unification scale, the inputs to the supergravity model
are a universal scalar mass, $M_0$, a common gaugino mass, $M_{1/2}$,
and a trilinear scalar coupling $A_0$.  As in the gauge-mediated model,
radiative breaking of electroweak symmetry breaking is imposed.

\begin{figure}[t]
\epsfysize=2in
\epsffile[-30 370 110 570]{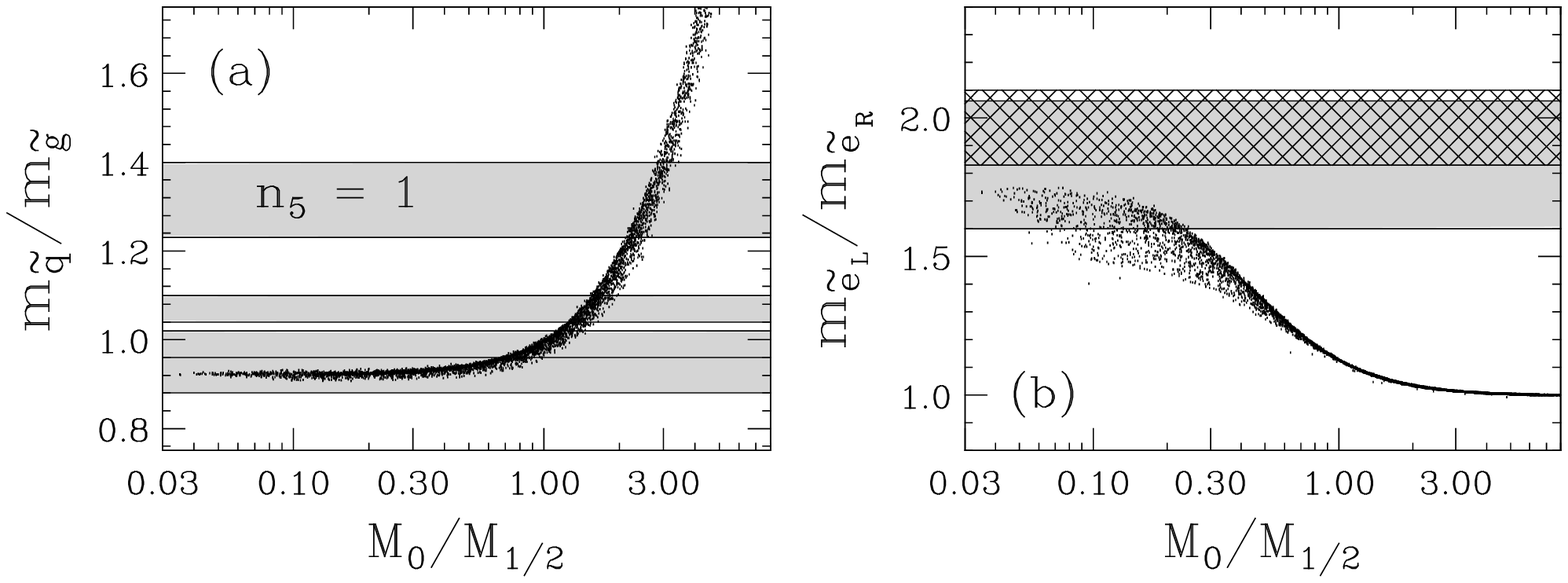}
\begin{center}
\parbox{5.5in}{
\caption[]{(a) The ratio of the (first or second generation) squark
mass to the gluino mass in the supergravity-inspired (dots) and
gauge-mediated models (bands).  The four bands in the gauge-mediated
case correspond to $n_5=1,2,3,4$. The largest ratio (uppermost band)
is for $n_5 = 1$.  (b) The ratio of the left-handed to the
right-handed selectron mass in the two models. The scatter plot shows
the supergravity model prediction. The bands mark, in the
gauge-mediated model with $n_5=4$, the regions where $m_{\tilde
e_R}<M_Z$ (grey) and $m_{\tilde e_R}>M_Z$ (hatched).
\label{f.compare}}}
\end{center}
\end{figure}

In general, the two models predict qualitatively different spectra.
The supergravity model includes separate scales for the gauginos and
the scalars, so the scalars can be much heavier than the gauginos.
The minimal gauge-mediated model has only one scale, so the ratios of
the gaugino to the scalar masses are more or less fixed.  We illustrate
this in Fig.~\ref{f.compare}(a), where we show the ratio of the (first
or second generation) squark mass to the gluino mass in the two
models.\footnote{The supergravity scatter plots were generated as in
Ref.~\cite{PBMZ}.}  In the supergravity case, this ratio varies from
about 0.9 to 5.  For gauge-mediated models, the ratio depends on
$n_5$.  For $n_5=1$, the squarks are 20 to 40\% heavier than the
gluinos, while for $n_5=4$ the squarks are about 5 to 10\% lighter.
(These numbers and the bands on the figure ignore the region $M \lesssim2
\Lambda$, where the ratio $m_{\tilde q}/m_{\tilde g}$ falls by 20\%.)
Note that gauge-mediated models with $n_5=4$ predict the same ratio of
squark to gluino mass as the supergravity models with $M_0 \lesssim
M_{1/2}$.

In the region where the two models predict the same gluino and squark
masses, they do not typically predict the same slepton masses.  We
illustrate this in Fig.~\ref{f.compare}(b), where we show a
supergravity scatter plot of the ratio of the left-handed to the
right-handed selectron mass, versus $M_0/M_{1/2}$.  On the scatter
plot we superimpose the prediction of the gauge-mediated model, for
$n_5=4$, divided up into regions of small and large right-handed
selectron mass ($m_{\tilde e_R}<M_Z$ and $m_{\tilde e_R}>M_Z$).  In
the small $m_{\tilde e_R}$ region, the $D$-term contributions to
the slepton masses are important. The $D$-term contributions can
make the selectron ratios coincide with those from the
supergravity model in the region $M_0 \ll M_{1/2}$ if $\tan\beta$
is less than 3 in supergravity model and larger than 5 in the
gauge-mediated model.

\begin{table}[t]
\begin{center}
\begin{tabular}{ccc|cccc}
\multicolumn{3}{c|}{$\Delta m / m < 2 \%$} &
\multicolumn{4}{  c }{$\Delta m / m > 2 \%$} \\ \hline
  & GMSB & SUGRA &      & GMSB & SUGRA & \% diff. \\ \hline
$m_{\tilde g}$ & 361 & 366 &
$m_h$ & 104 & 77 & 26  \\
$m_{\tilde\chi_1^+}$ & 83 & 82 &
$m_{\tilde\chi_2^+}$ & 212 & 303 & $-43$  \\
$m_{\tilde\chi_2^0}$ & 87 & 87 &
$m_{\tilde \chi_1^0}$ & 49 & 45 & 9 \\
$m_{\tilde e_L}$ & 107 & 106 &
$m_{\tilde\chi_3^0}$ & 175 & 276 & $-58$  \\
$m_{\tilde e_R}$ & 64 & 64 &
$m_{\tilde\chi_4^0}$ & 210 & 307 & $-46$  \\
$m_{\tilde u_L}$ & 336 & 337 &
$m_{\tilde\nu_e}$ & 72 & 85 & $-19$  \\
$m_{\tilde  u_R}$ & 328 & 329 &
$m_{\tilde\nu_\tau}$ & 71 & 85 & $- 19$  \\
$m_{\tilde d_L}$ & 345 & 343 &
$m_{\tilde\tau_L}$ & 115 & 105 & 8 \\
$m_{\tilde d_R}$ & 330 & 329 &
$m_{\tilde\tau_R}$ & 46 & 63 & $-37$  \\
$m_{\tilde t_L}$ & 382 & 382 &
$m_{\tilde t_R}$ & 281 & 235 & 16  \\
$m_{\tilde b_L}$ & 309 & 304 &
$m_{\tilde b_R}$ & 330 & 322 & 3    \\
&& & $m_A$ & 170 & 318 & $-87$   \\
\end{tabular}
\parbox{5.5in}{
\caption[]{Comparison of the spectra of a gauge-mediated (GMSB) and
supergravity-inspired (SUGRA) model. The masses are in units of GeV.}}
\end{center}
\end{table}

If we go even further, we find that the two models can predict the
same gluino, squark, charged slepton and light chargino masses. We
give an example in Table~1. This table lists the spectra for the
supergravity model parameters $\tan\beta=2.1$, $M_0=5$ GeV,
$M_{1/2}=145$ GeV, $A_0=686$ GeV, and $\mu<0$, and for the
gauge-mediated model parameters $\tan\beta=14$, $\Lambda=11$ TeV,
$M/\Lambda=3716$, $n_5=4$ and $\mu<0$.  The overlap can occur only for
small selectron and $\tilde\chi_1^0$ masses, where the slepton
$D$-terms and the gaugino/Higgsino mixing come into play.

Table 1 also illustrates that the models differ in their predictions
for other observables.  If we scan over the parameter spaces of the
two models and match the gluino masses to 8\%, and the light chargino
and (first and second generation) squark and selectron masses to 2\%,
we find that the two models predict different values of $\tan\beta$,
as well as significantly different masses for the neutralinos, the
sneutrinos, the heavy chargino and the light Higgs boson.  They also
predict different values for the Higgs parameters $\mu$ and $m_A$,
and, typically, different masses for the third-generation squarks and
sleptons.

\begin{figure}[t]
\epsfysize=2in
\epsffile[0 358 140 548]{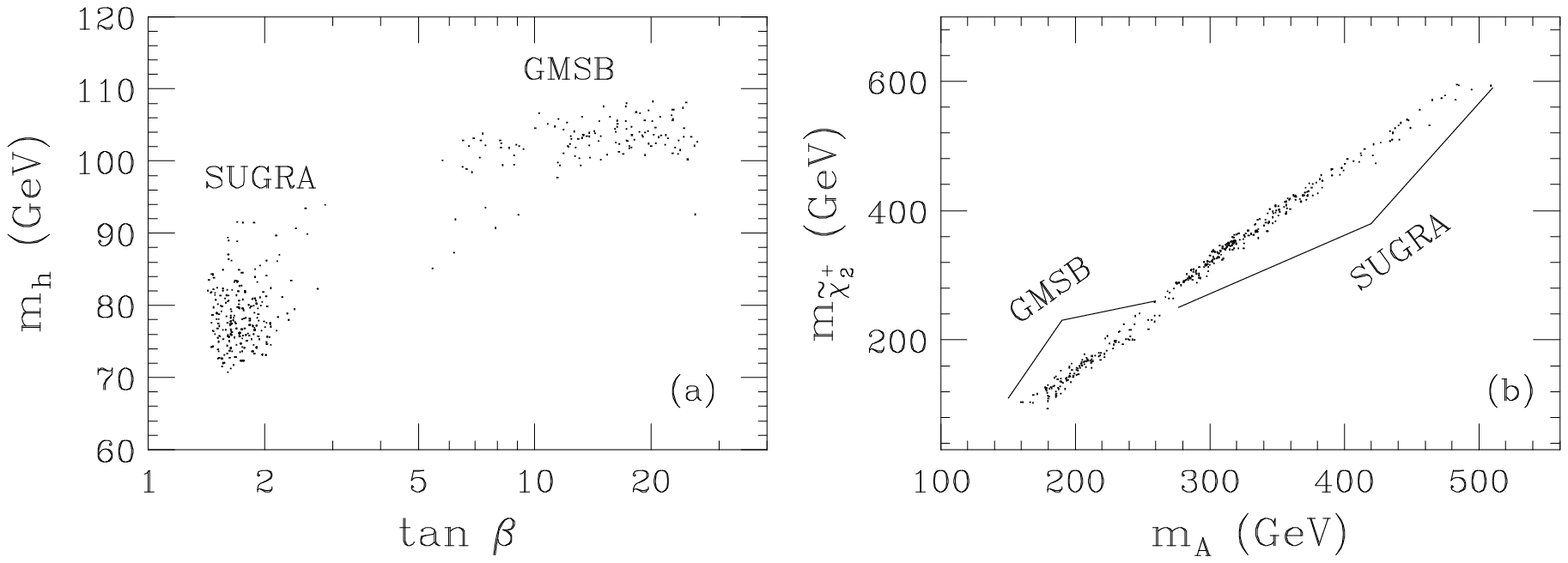}
\begin{center}
\parbox{5.5in}{
\caption[]{Scatter plots of observables in the
regions of parameter space where the gauge-mediated
and supergravity models predict the same gluino,
squark, slepton and light chargino masses (see text).
Figure (a) shows a scatter plot of $m_h$ versus
$\tan\beta$ in the gauge-mediated (GMSB)
and supergravity models (SUGRA).  Figure (b) shows
$m_{\tilde\chi_2^+}$ versus $m_A$.
\label{f.comp}}}
\end{center}
\end{figure}

We illustrate these differences in Fig.~\ref{f.comp}, where we show
scatter plots of the values of $\tan\beta, m_h, m_A,$ and
$m_{\tilde\chi_2^+}$ in the two models.  We see that the models are
clearly distinguished.  Note that since both models have essentially
one scale, there is a strong correlation between the heavy Higgs and
Higgsino masses. (The two scales are $\Lambda$ and $M_{1/2}$, since
$M_0\simeq0$.)  We conclude that it is possible, but not probable,
that measurements of the gluino, squark, slepton and light chargino
masses may not be enough to discriminate the gauge-mediated from the
supergravity model.  However, additional measurements would easily
rule out one (or both) of the models.

\section{Properties of the NLSP}

In the preceding section, we saw that the NLSP is either the
light tau slepton, $\tilde\tau_1$, or the lightest neutralino,
$\tilde\chi_1^0$.  Over most of the parameter space, the number of
$5+\overline{5}$ pairs determines the NLSP.  For $n_5 = 1$, the
$\tilde\chi_1^0$ tends to be the NLSP.  For larger $n_5$, the balance
tips towards the $\tilde\tau_1$ because as $n_5$ increases, the scalar
masses increase less than their gaugino counterparts.

\begin{figure}[t]
\epsfysize=2.5in
\epsffile[-180 180 -40 545]{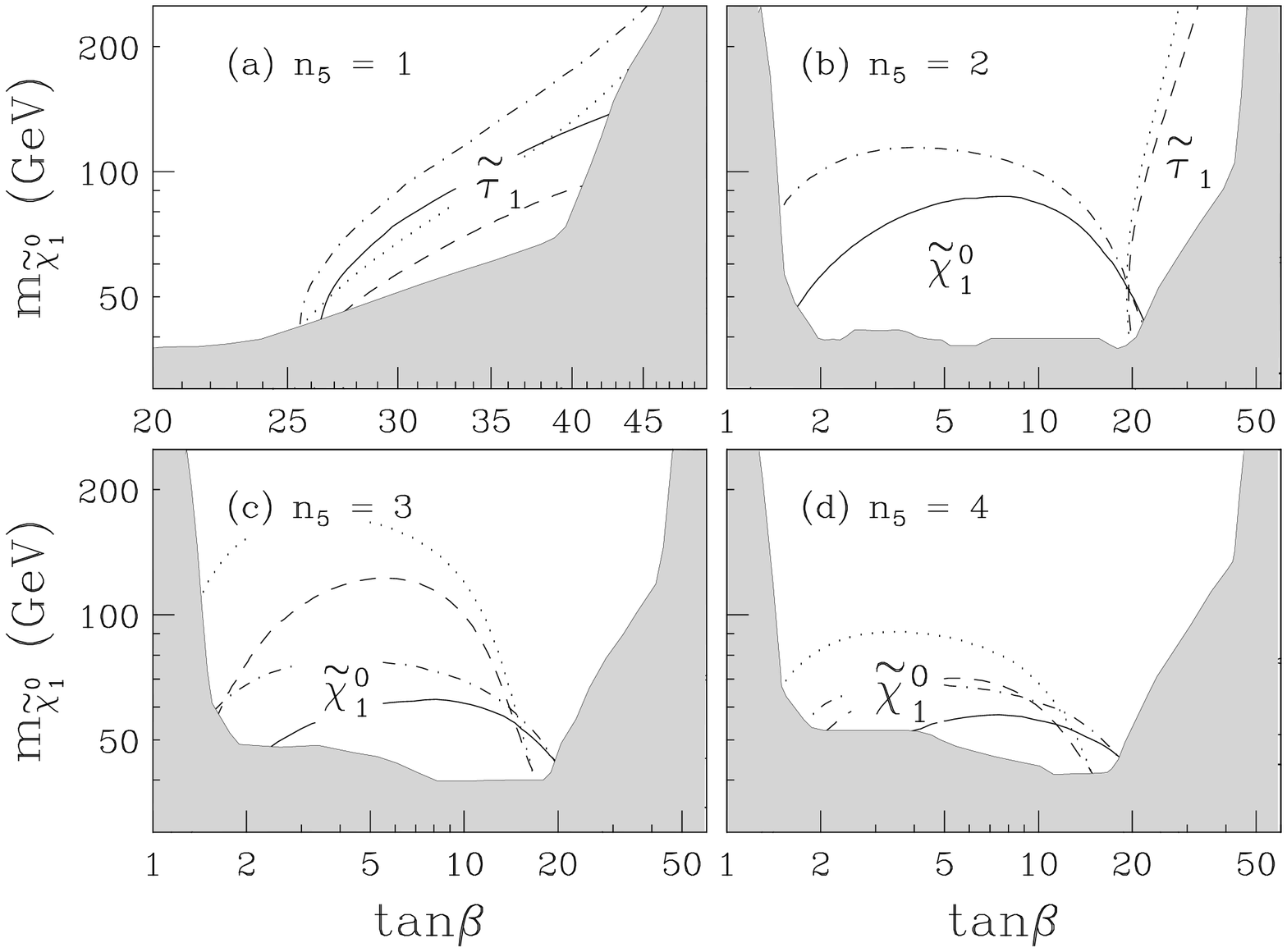}
\begin{center}
\parbox{5.5in}{
\caption[]{Contours marking the boundary where the mass of the light
tau slepton is equal to the mass of the lightest neutralino, for $n_5
= 1, 2, 3$ and $4$.  The solid (dashed) line corresponds to $\mu >
0$, $M/\Lambda = 2$ $(10^4)$.  The dot-dashed (dotted) line
corresponds to $\mu < 0$, $M/\Lambda = 2$ $(10^4)$.  The excluded
region is plotted for $\mu > 0$ and $M/\Lambda = 2$.  The label
$\tilde\chi_1^0$ ($\tilde\tau_1$) marks the region where the
neutralino (tau slepton) is the NLSP.
\label{f.Xtau}}}
\end{center}
\end{figure}

In Fig.~\ref{f.Xtau} we plot contours of $m_{\tilde\chi_1^0} =
m_{\tilde\tau_1}$ in the $(m_{\tilde\chi_1^0},\tan\beta)$ plane, for
each value of $n_5$ between 1 and 4.  Each subfigure contains four
contours, two for each sign of $\mu$, and two for $M/\Lambda=2$ and
$10^4$.  We see that for $n_5=1$, the neutralino is the NLSP, except
for small regions at large $\tan\beta \gtrsim 25$ and small
$m_{\tilde\chi_1^0}$.  For $n_5=2$ and $M/\Lambda = 10^4$, $\chi_1^0$
is the NLSP for $\tan\beta \lesssim 20-30$.  For $n_5=2$ and $M/\Lambda =
2$, the light tau slepton is the NLSP over most of the parameter
space, as it is for $n_5=3$ or 4 and arbitrary $M/\Lambda$.  For $n_5
> 2$, the neutralino is the NLSP at very small $m_{\tilde\chi_1^0}
\lesssim 100$ GeV.

These results are subject to significant uncertainties from the
unknown messenger-scale thresholds.  In particular, the line where
$m_{\tilde\chi_1^0} = m_{\tilde\tau_1}$ is sensitive to potentially
large one-loop corrections to the mass of the light tau slepton.  We
find that 5\% variations in the messenger-scale boundary conditions
give rise to a $m_{\tilde\chi_1^0} - m_{\tilde\tau_1}$ mass difference
of up to 30 GeV in the region where the difference is less than 100
GeV.

The collider phenomenology in gauge-mediated models depends crucially
on the nature of the NLSP, especially when it decays inside the
detector.  The lifetime of the NLSP depends on its mass, $m$, and on
the ratio $M/\Lambda$, as follows,
\begin{equation}
\tau \propto {F_S^2\over m^5}
 \propto {1\over m}\left({M\over\lambda\Lambda}\right)^2  .
\label{ctau}
\end{equation}
For the case at hand, where the NLSP is either the $B$-ino or the
$\tilde\tau_1$, we can put in the appropriate factors of $\pi$ and
$\alpha_1$ and write the lifetime in units of meters,
\begin{equation}
c\tau \simeq \left({100~{\rm GeV}\over m}\right)
\left({M\over\lambda\Lambda}\right)^2 \times 10^{-5}~\rm{meters} .
\label{length}
\end{equation}
For $\beta\gamma = \sqrt{E^2/m^2 - 1} \simeq 1$, this is the
approximate decay length.  We see that for $\beta\gamma \simeq 1$,
$\lambda \simeq 1$ and $M/\Lambda \lesssim 100$, the NLSP will decay
inside the detector.  However, for sufficiently small $\lambda$ it
will decay outside the detector.

\begin{figure}[t]
\epsfysize=2in
\epsffile[-10 390 130 580]{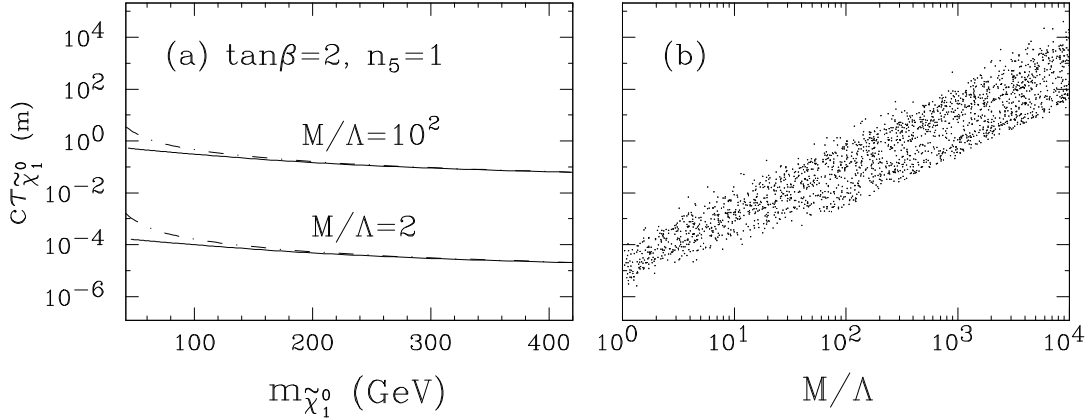}
\begin{center}
\parbox{5.5in}{
\caption[]{(a) The lifetime of the NLSP, for $\tan\beta=2$ and $n_5
=1$, with $\mu>0$ (solid line) or $\mu<0$ (dot-dashed).  For fixed
$m_{\chi_1^0}$, the lifetime scales like $(M/\Lambda)^2$.  (b) A
scatter plot of the neutralino lifetime, restricted to cases where the
neutralino is the NLSP.
\label{f.Xlife}}}
\end{center}
\end{figure}

In Fig.~\ref{f.Xlife} we illustrate the neutralino lifetime in
different regions of parameter space.  In Fig.~\ref{f.Xlife}(a) we
plot the lifetime versus the neutralino mass, for $M/\Lambda=2$ and
100, with $n_5=1$, $\lambda=1$ and $\tan\beta=2$.  In
Fig.~\ref{f.Xlife}(b) we show a scatter plot of the neutralino
lifetime versus $M/\Lambda$.  The lifetime increases by eight orders
of magnitude as we increase $M/\Lambda$ from 1 to $10^4$.  In this and
the next scatter plot, we vary the parameters $n_5$ from 1 to 4,
$\Lambda$ from 16 TeV$/n_5$ to 300 TeV$/\sqrt{n_5}$, $M/\Lambda$ from
1.03 to $10^4$, and $\tan\beta$ from 1.2 to 70.  The latter three
variables are sampled on logarithmic measures, subject to the
phenomenological constraints discussed in sect. 2.

\begin{figure}[t]
\epsfysize=2.5in
\epsffile[-180 180 -40 545]{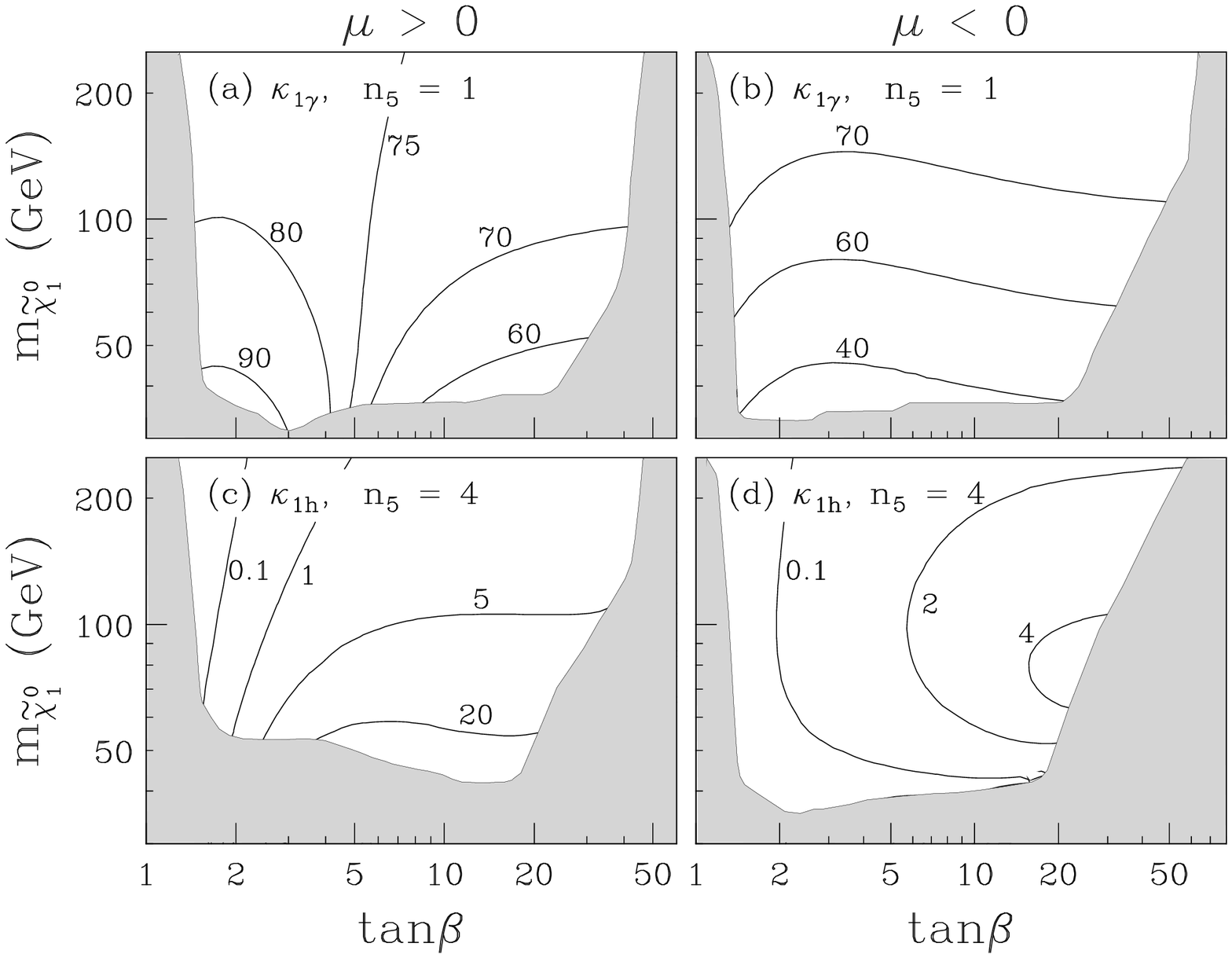}
\begin{center}
\parbox{5.5in}{
\caption[]{Figures (a) and (b) [(c) and (d)] show the photino
component of the lightest neutralino, $\kappa_{1\gamma}$ [the
Higgsino component, $\kappa_{1h}$], in per cent, for
$M/\Lambda = 2$ and $n_5 = 1$ [$n_5 = 4$].
\label{f.ki}}}
\end{center}
\end{figure}

If the tau slepton is the NLSP, it will decay to a $\tau$ lepton and a
gravitino with a branching fraction that is essentially 100\%.  The
lightest neutralino has many more decay modes because it has Higgsino
and gaugino components.  The $\tilde\chi_1^0$ can decay to either a
Higgs or a gauge boson, plus a gravitino, $\tilde G$.  In
Fig.~\ref{f.ki} we show the Higgsino and photino components of the
$\tilde\chi_1^0$.  In Figs.~\ref{f.ki}(a) and \ref{f.ki}(b) we set
$n_5=1$ and see that the NLSP is approximately 60 to 90\% (40\% to
70\%) photino for $\mu > 0$ $(\mu < 0)$.  The Higgsino component is
quite small.  The Higgsino component can be larger if $n_5$ is larger,
because $|\mu|$ is reduced relative to the gaugino masses.  In
Fig.~\ref{f.ki}(c) we see that for $n_5=4$ and $\mu>0$, the Higgsino
component of the NLSP can be large. In fact, it can be as large as
50\%.

The $\tilde \chi_i^0$ partial widths can be readily computed.  We express
them in units of
\begin{equation}
{\cal A} = {m_{\tilde\chi^0_i}^5\over96\pi M_{\tilde G}^2M_P^2} =
{m_{\tilde\chi^0_i}^5\over32\pi F_S^2} .
\end{equation}
We first write the well-known decay rate to a photon and the gravitino,
\begin{equation}
{\cal A}^{-1}\Gamma(\tilde\chi_i^0\rightarrow\tilde G\gamma)
 =  2 \kappa_{i\gamma}.
\label{phot}
\end{equation}
For the remaining decay modes, we give the three body decay formula to
account for the decay to standard model fermions via real or virtual
boson exchange.  If we sum over final-state standard-model fermions,
we find
\begin{eqnarray}
{\cal A}^{-1} \Gamma(\tilde\chi_i^0\rightarrow\tilde Gf^+f^-)
&=& 2 \kappa_{iZ_T} I_1(Z) + \kappa_{iZ_L} I_0(Z) \\[2mm]
&+&80s_Wc_W\biggl[{3-8s_W^2
\over63-120s_W^2+160s_W^4}\biggr]\kappa_{iZ\gamma}
\left(I_1(Z)-I_0(Z)\right)\nonumber\\[2mm]
&+& \kappa_{i\gamma}{2\alpha\over3\pi}\sum_f N_c^f e_f^2
\left(-{25\over12}
-\ln {s^f_{\rm min}\over m_{\tilde\chi_i^0}^2}\right)\nonumber\\
&+& \sum_{\varphi=h,H,A} {\rm BR}(\varphi\rightarrow f^+f^-)
\kappa_{i\varphi} I_1(\varphi)\nonumber\\
&-& 2 \Biggl[{\rm BR}(h\rightarrow f^+f^-){\rm BR}(H\rightarrow f^+f^-)
\Biggr]^{1\over2} \kappa_{iHh} I_{10}(h,H) ,\nonumber
\end{eqnarray}
where the sum $\sum_f$ is over all fermions with mass less than
$m_{\tilde\chi_i^0}/2$, and $N_c^f$ is the number of colors: 3 for
quarks and 1 for leptons.  In the virtual photon contribution, $s^f_{\rm
min}$ is a detector dependent cut-off on the fermion pair invariant
mass. Also, we define
\begin{eqnarray}
\kappa_{i\gamma} &=& |N_{i1}c_W+N_{i2}s_W|^2 ,\\
\kappa_{iZ_T} &=& |N_{i2}c_W-N_{i1}s_W|^2 ,\nonumber\\
\kappa_{iZ\gamma} &=& c_Ws_W\left(|N_{i2}|^2-|N_{i1}|^2\right)
         + (c_W^2-s_W^2) {\cal R}e \left(N_{i1}^*N_{i2}\right) ,\nonumber\\
\kappa_{iZ_L} &=& |N_{i3}c_\beta-N_{i4}s_\beta|^2 ,\nonumber\\
\kappa_{iA} &=& |N_{i4}c_\beta+N_{i3}s_\beta|^2 ,\nonumber\\
\kappa_{iH} &=& |N_{i3}c_\alpha+N_{i4}s_\alpha|^2 ,\nonumber\\
\kappa_{ih} &=& |N_{i4}c_\alpha-N_{i3}s_\alpha|^2 ,\nonumber\\
\kappa_{iHh} &=& s_\alpha c_\alpha\left(|N_{i4}|^2-|N_{i3}|^2\right)+
(c_\alpha^2-s_\alpha^2) {\cal R}e \left(N_{i3}^*N_{i4}\right) .\nonumber
\end{eqnarray}
In writing the $H$-$h$ interference contribution, we have neglected
all Yukawa couplings except those of the the bottom and tau.  The
$\gamma$-$Z$ ($H$-$h$) interference terms go to zero in the limit
$m_{\tilde\chi_i^0}\gg M_Z$ ($m_{\tilde\chi_i^0}\gg m_H$).  Note that
if $m_{\tilde\chi_i^0} > M_\varphi$ ($\varphi = h,H,A$), the decay rate
${\cal A}^{-1} \Gamma(\tilde\chi_i^0\rightarrow\tilde G\varphi) =
\kappa_{i \varphi} I_1(\varphi)$.

In these expressions, the integrals $I_n$ are given by
\begin{equation}
I_n(\varphi) = {\epsilon_\varphi\over\pi}\int_0^1 dx
 {(1-x)^4(x/R_\varphi)^n\over(x-R_\varphi)^2+\epsilon_\varphi^2} ,
\end{equation}
with $R_\varphi=M_\varphi^2/m_{\tilde\chi_i^0}^2$ and
$\epsilon_\varphi=\Gamma_\varphi M_\varphi/m_{\tilde\chi_i^0}^2$.
These integrals reduce to $(1-R_\varphi)^4$ for small
$\epsilon_\varphi$ (i.e. in the narrow width approximation).  The
integral $I_{10}(h,H)$ is
\begin{equation}
I_{10}(h,H) = {1\over\pi}
\sqrt{{\epsilon_h\epsilon_H\over R_h R_H}}\int_0^1 dx
{x (1-x)^4 \biggl[\left(x-R_H\right)\left(x-R_h\right)
+ \epsilon_h\epsilon_H\biggr]\over
\left((x-R_h)^2+\epsilon_h^2\right)
\left((x-R_H)^2+\epsilon_H^2\right)} .
\end{equation}

Our results complete the formulae for the two-body decay rates of the
$\chi_i^0$ given in ref.~\cite{Martin et al}.  Our results include the
contribution of the virtual photon and the contributions from
$Z$-$\gamma$ and $H$-$h$ interference.  Note that the formula for
$\Gamma( \tilde \chi_i^0 \rightarrow \tilde G +$ Higgs boson)
contains the function $I_1$, while the formula for $\Gamma( \tilde
\chi_i^0 \rightarrow \tilde G +$ longitudinal Z boson) contains the
function $I_0$.  At first sight one might think that this violates the
electroweak equivalence theorem, since the decay rate to the
longitudinal $Z$ is not equal to the decay rate to the Goldstone
boson. However, the equivalence principle does indeed hold in the
applicable regime, since the two functions approach each other for
$m_{\tilde\chi_i^0}\gg M_Z$.

\begin{figure}[t]
\epsfysize=2in
\epsffile[-20 410 120 600]{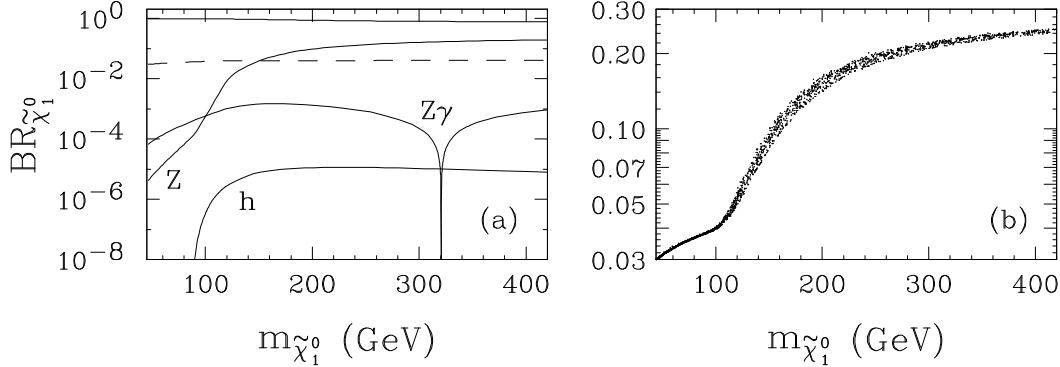}
\begin{center}
\parbox{5.5in}{
\caption[]{(a) The branching ratios of the $\tilde\chi^0_1$ in the
case $M/\Lambda = 2$, $n_5 = 1$ and $\mu > 0$. The solid line near 1
is the branching ratio to a photon and the gravitino, and the dashed
line is the branching ratio to fermions via a virtual photon. (b) A
scatter plot of the neutralino branching fraction to fermion pairs, in
the cases where the neutralino is the NLSP.
\label{f.BR}}}
\end{center}
\end{figure}

In Fig.~\ref{f.BR}(a) we show the branching fractions of the
$\tilde\chi_1^0$ versus $m_{\tilde\chi_1^0}$, for $\tan\beta=2$,
$n_5=1,$ and $M=2\Lambda$.  (We plot the magnitude of the $Z$-$\gamma$
interference contribution, $|\Gamma_{Z\gamma}| /\Gamma_{\rm tot}$, and
we take into account 5 quark flavors.)  The decay to the photon
dominates, as expected, since the $\tilde\chi_1^0$ has a large photino
component.  The branching fraction to fermions via an off-shell photon
varies from about 3 to 4\% for $s^f_{\rm min}=1$ GeV.

In the parameter space associated with the minimal models, the
branching ratio to fermion pairs via heavy Higgs boson exchange is
negligible.  For the parameters corresponding to Fig.~\ref{f.BR}, the
heavy Higgs exchange branching ratio is of order $10^{-12}$. The
branching ratio associated with $H$-$h$ interference and the branching
fraction associated with virtual $h$-boson exchange are about an order
of magnitude larger.

Because of the kinematical suppression, the branching fraction to the
$Z$-boson rises slowly above threshold, to about 15\% (20\%) for
$m_{\tilde\chi_1^0} = 2M_Z$ $(3M_Z)$.  We illustrate this in
Fig.~\ref{f.BR}(b), where we show a scatter plot of the branching
fraction of the lightest neutralino to a gravitino and a fermion pair.
We see that the branching ratio depends only on the neutralino mass.
This would not be true if the part of parameter space in which the
$\tilde\chi_1^0$ has a large Higgsino component ($n_5=4, \mu>0, $small
$\Lambda$, see Fig.~\ref{f.ki}(c)) led to a sizable branching ratio to
the Higgs boson.  However, in this region of parameter space either
the $\tilde\tau_1$ is the NLSP or the $\tilde\chi_1^0$ mass is below
the $h$-boson threshold.  Hence, the associated partial width is
negligible.

\begin{figure}[t]
\epsfysize=2.5in
\epsffile[-140 220 0 555]{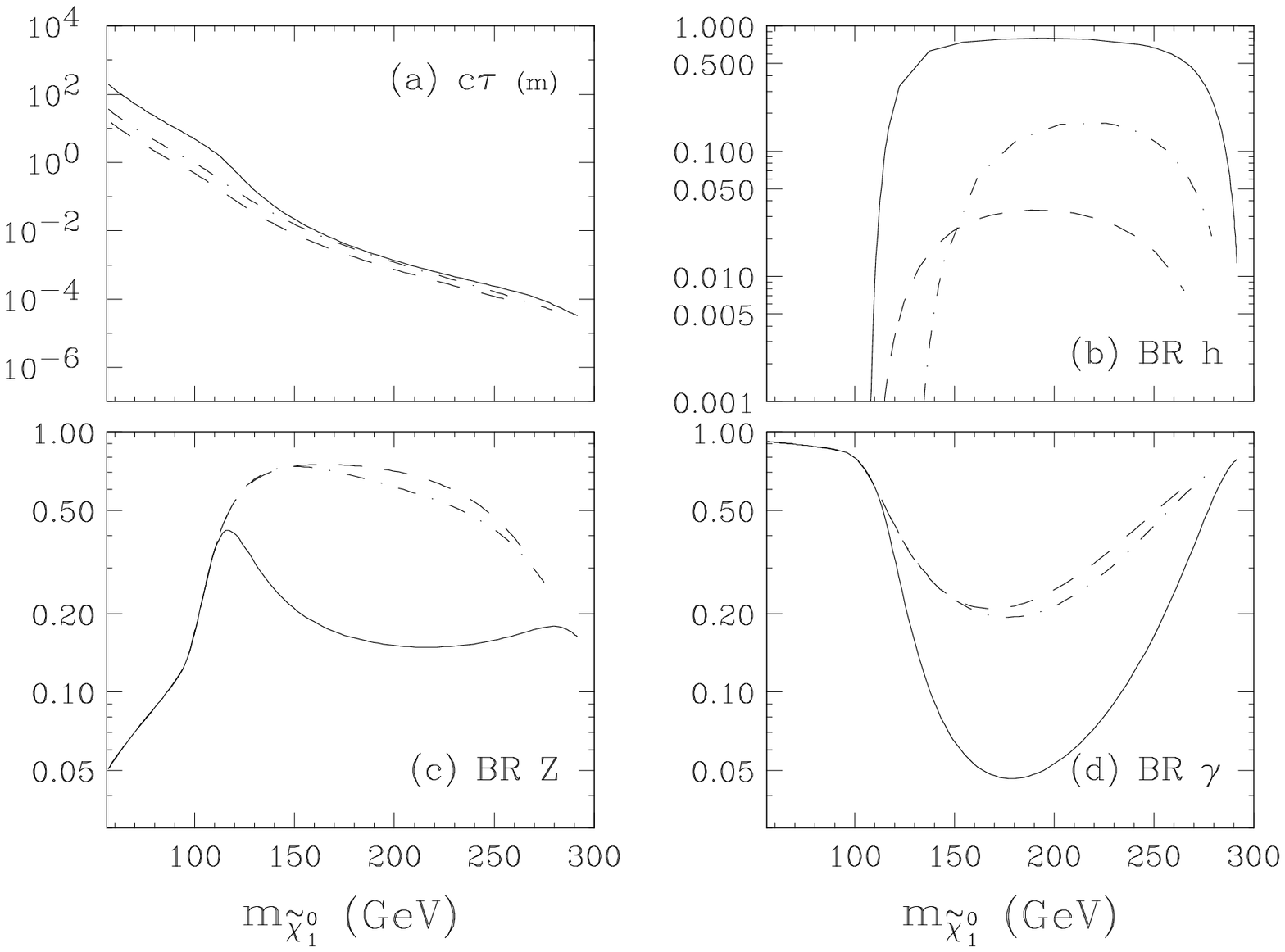}
\begin{center}
\parbox{5.5in}{
\caption[]{The (a) lifetime and (b-d) branching fractions of the
NLSP, for the case where the $\tilde\chi^0_1$ is primarily Higgsino,
with $n_5 = 1$, $\lambda=1$ and $\Lambda = 200$ TeV.  The
$m_{\tilde\chi_1^0}$ values are obtained by varying $\mu$
from 50 to 350 GeV. The solid
(dashed) lines are for $\tan\beta=2$, with $\mu$ positive (negative).
The dot-dashed lines correspond to $\tan\beta=40$, $\mu>0$.
(The curves for $\tan\beta=40$, $\mu<0$ are almost identical.)
\label{f.Xhiggs}}}
\end{center}
\end{figure}

As discussed above, the mechanism that generates $B(M)$ and $\mu(M)$
might result in extra terms to the scalar masses beyond those in
eq.~(\ref{bc2}).  In this case, the NLSP can decay predominately
into a Higgs boson.  To study this possibility, we take $\mu$
and $B$ to be independent low-energy parameters, and focus our
attention on the case of small $\mu$, so the lightest neutralino is
predominantly Higgsino.  In Fig.~\ref{f.Xhiggs} we plot the lifetime
and branching fractions of the $\tilde\chi_1^0$ as a function of
$m_{\tilde\chi_1^0}$, for various choices of parameters.  Note that in
the region $m_{\tilde\chi_1^0}\lesssim250$ GeV the lightest neutralino is
largely Higgsino, and $m_{\tilde\chi_1^0} \simeq |\mu|$.

The lifetime and branching fractions of the NLSP in the Higgsino
region are shown in Fig.~\ref{f.Xhiggs}.  In Fig.~\ref{f.Xhiggs}(a) we
see that the lifetime varies by 6 or 7 orders of magnitude as
$m_{\tilde\chi^0_1}$ varies over less than one order of magnitude.
This can be contrasted with the case where the neutralino is
dominantly $B$-ino, where the lifetime varies by less than one order
of magnitude (see eq.~(\ref{ctau}) and Fig.~\ref{f.Xlife}).  The
branching fractions of the lightest neutralino are shown in
Figs.~\ref{f.Xhiggs}(b-d).

These results can be understood as follows.  In the small $\mu$ region
the NLSP is primarily Higgsino, so it prefers to decay to a Higgs
boson.  Below $h$ and $Z$ threshold, however, it is forced to decay
through its suppressed photino component.  This accounts for the long
lifetime and the large photino branching fraction for small
$m_{\tilde\chi^0_1}$.  As the mass increases, the $h$ and $Z$ channels
open, so the lifetime decreases and the branching fractions are
determined by $\kappa_{1 Z_L}$ and $\kappa_{1h}$.  For small
$\tan\beta$, the decay to $h$ ($Z$) dominates for $\mu$ positive
(negative).  For large $\tan\beta$, $\kappa_{1 Z_L} \simeq
\kappa_{1h}$ for each sign of $\mu$.  In this case the branching
fraction to $Z$ is larger because the Higgs is heavier, so $I_0(Z) >
I_1(h)$.  Finally, as $m_{\tilde\chi_1^0}$ increases still further,
the decay to photons again dominates, because in this region, the NLSP
is predominately photino.  For $m_{\tilde\chi_1^0} \simeq 300$ GeV,
the decay length approaches $10^{-5}$ meters.

The neutralino lifetime in the Higgsino region can be readily scaled
for other values of $n_5$ and $M/\Lambda$.  As before, for fixed
$m_{\tilde\chi_1^0}$, $c\tau$ scales like $(M/\Lambda)^2$.
For fixed $M/\Lambda$ and fixed gaugino masses, it scales like
$1/n_5^4$  because $\tau \propto F^2_S/\mu^5 \propto \Lambda^4/\mu^5$.

These results should be contrasted with those of the minimal models,
where we found that, above $h$ threshold, the branching ratio to the
$h$ varies from $10^{-8}$ to $10^{-4}$.  Therefore collider events
with four $b$-jets and missing energy would imply a non-minimal Higgs
sector in these models.

\section{Conclusions}

In this paper we examined the detailed low-energy spectrum of
gauge-mediated supersymmetry breaking models.  We used two-loop
renormalization group equations for the gauge and Yukawa couplings,
and for the soft supersymmetry breaking parameters.  We imposed
consistent one-loop radiative electroweak symmetry breaking under
the assumption that the mechanism that generates the $\mu$ and $B$
parameters does not induce any extra contributions to the scalar
masses.

We examined the phenomenology in the case of an arbitrary number of
$5+\overline{5}$ messenger fields, and in the case that the messenger
scale $M$ is greater than $\Lambda$.  We began by examining
electroweak symmetry breaking.  We initially considered the minimal
case, $n_5=1$ and $M=2 \Lambda$.  In the parameter space we
considered, with $\Lambda < 300$ TeV, we found that $|\mu(M)|$
varies from 150 GeV to over 1 TeV.  We also found that $|B(M)/\mu(M)|$
varies from near (but not equal to) zero to about 500 GeV.
We found the phenomenologically
interesting region $B(M) \simeq 0$ to be compatible with electroweak
symmetry breaking in the region of large $\tan\beta$, for either sign
of $\mu$.  For larger $M/\Lambda$, we found $B(M) \simeq 0 $ requires
$\mu < 0$.

We examined the spectrum and illustrated how it depends on $n_5$
and $M/\Lambda$.  We found that the spectrum is not qualitatively
affected by an increase in $M/\Lambda$.  For $n_5>1$, we found that
the light tau slepton is the NLSP over most of the parameter
space.  It decays to a tau lepton and a gravitino.

By varying the boundary conditions of the soft parameters at the
messenger scale by $5$\%, we determined the sensitivity of the
supersymmetric spectrum to higher-order messenger-sector corrections.
We found that in the region $B(M) \simeq 0$, $B(M)$ varies by $10$
GeV.  Furthermore, the spectrum varies by $5\%$, except for the light
Higgs mass, which is essentially unchanged, and the third-generation
squark and slepton masses.  In the most extreme case, we found a
$70\%$ variation in the light tau slepton mass, which implies
a substantial uncertainty in the identification of the NLSP.

We compared the predictions for the spectra in the
gauge-mediated and supergravity-inspired models.   In general, the
two models predict qualitatively different spectra.  We found that it
is possible for the two models to give the same gluino, light
chargino, and (first and second generation) squark and slepton masses.
Such a match is only possible for very light masses, where $D$-terms
and gaugino/Higgsino mixing is important.  In this region, the models
can be distinguished by their predictions for other observables,
such as $\tan\beta$, as well as the sneutrino, neutralino, heavy
chargino, and Higgs boson masses.

For $n_5=1$, we found that the $\tilde\chi_1^0$ is usually the NLSP.
The $\tilde\chi_1^0$ decays to the gravitino and either an
(on- or off-shell) gauge or Higgs boson.  We derived the neutralino
decay rate and branching fractions.  The lightest neutralino tends to
be gaugino-like, and decays to a photon and a gravitino.  However,
in the region $m_{\tilde\chi_1^0} \gtrsim 100$ GeV, the branching fraction
to the $Z$ can be larger than 20\%.  Leptons are easier to track than
photons, so this mode has the potential to permit a precise neutralino
lifetime measurement. We found that for $n_5=4$ the Higgsino component
of the $\tilde\chi_1^0$ can be as large as 50\%. Nevertheless, over
the entire parameter space the branching fraction to the light Higgs
boson is less than $10^{-4}$.

We also examined the branching fraction of the neutralino in the
Higgsino region.  In this case the branching fraction to the on-shell
$h$-boson can reach over 80\% well above threshold.  Since this Higgsino
region can only occur for models with non-minimal Higgs sectors, the
observation of 4 $b$-jets plus missing energy would be an important
step towards understanding the origin of the $\mu$ term.

Models with gauge-mediated supersymmetry breaking offer the appealing
possibility that the origin of supersymmetry breaking is
experimentally accessible.  Once supersymmetry is discovered, detailed
study of the supersymmetric particles, along the lines suggested here,
might well prove to be the first step towards uncovering the mechanism
of supersymmetry breaking.

\section*{Acknowledgements}
One of us (D.M.P.) thanks S. Thomas and J. Wells for useful
conversations.

\end{document}